\documentclass[aps,pre, reprint, showpacs,superscriptaddress, nofootinbib]{revtex4-1}

\usepackage[english]{babel}
\usepackage{amsmath}
\usepackage{graphicx}
\usepackage{setspace}
\usepackage{bm}
\usepackage{color}
\usepackage{hyperref}

\definecolor{hreflinkcolor}{rgb}{0.13,0.17,0.83}
\hypersetup{colorlinks=true,urlcolor=hreflinkcolor,
        linkcolor=hreflinkcolor,citecolor=hreflinkcolor}

\newcommand{\Changes}[1]{#1}
\newcommand{\ESedit}[1]{#1}

\newcommand{\refeq}  [1] {Eq.~(\ref{#1})}

\begin{document}

\title{Relativistic breather-like  solitary waves with linear polarization in cold plasmas }

\author{G. S\'anchez-Arriaga}
\affiliation{Departamento de F\'isica Aplicada, Escuela T\'ecnica Superior
de Ingenieros Aeron\'auticos, Universidad Polit\'ecnica de Madrid, Madrid, Spain}

\author{E. Siminos}
\affiliation{Max Planck Institute for the Physics of Complex Systems, N\"{o}thnitzer Str. 38,
D-01187 Dresden, Germany}

\author{V. Saxena}
\affiliation{Centre for Free-Electron Laser Science, Deutsches Elektronen-Synchrotron,
Notkestrasse 85, 22607 Hamburg, Germany}

\author{I. Kourakis}
\affiliation{Centre for Plasma Physics, School of Mathematics and Physics,
Queen's University Belfast, Belfast BT7 1NN, Northern Ireland, United Kingdom}

\begin{abstract}

Linearly polarized solitary waves, arising from the interaction of
an intense laser pulse with a plasma, are investigated. New
localized structures, in the form of exact \Changes{numerical}
nonlinear solutions of the one-dimensional Maxwell-fluid model for a
cold plasma with fixed ions are presented. Unlike stationary
circularly polarized solitary waves, the linear polarization gives
rise to a breather-like behavior and a periodic exchange of
electromagnetic energy and electron kinetic energy at twice the
frequency of the wave. A numerical method based on a
finite-differences scheme allows to compute a branch of solutions
within the frequency range $\Omega_{min}<\Omega<\omega_{pe}$, where
$\omega_{pe}$ and $\Omega_{min}$ are the electron plasma frequency
and   the frequency value for which the plasma density vanishes
locally, respectively. A detailed description of the spatio-temporal
structure of the waves and their main properties as a function of
$\Omega$ are presented. Small amplitude oscillations appearing in
the tail of the solitary waves, a consequence of the linear
polarization and harmonic excitation, are explained with the aid of
the Akhiezer-Polovin system. Direct numerical simulations of the
Maxwell-fluid model show that these solitary waves  propagate
without change for a long time.

\end{abstract}


\maketitle

\section{Introduction}

Electromagnetic relativistic solitary waves, commonly named
relativistic solitons, are self-trapped localized structures that
are  excited  during the interaction of a high-intensity laser with
a plasma. They typically consist on a region of rarified plasma
where a high amplitude electromagnetic wave, with frequency below
the laser frequency, is trapped. A solitary wave can be classified
according to its polarization (linear or circular), its group
velocity (standing or moving waves) and also its state (isolated
structure or embedded in long laser pulses). Since part of the laser
energy is dedicated to excite them, they may affect to certain laser
applications like inertial confinement fusion, particle
acceleration, plasma lens for ultra-intense laser focusing and
high-brightness X- and gamma rays generation \cite{Eliezer_08}. They
also offer an excellent opportunity to confront theoretical and
experimental works and enhance our knowledge on non-linear waves in
plasmas.

This great variety of solitary waves has been the subject of an
important research activity on plasma physics, including analytical
work, numerical simulations and laboratory experiments. Pioneer
works \cite{Akhiezer_56,Kozlov_79,Kaw_92} were followed by an
intense activity mainly in the framework of the relativistic fluid
model. Exact 1-dimensional circularly polarized solitary waves
solutions, including isolated
\cite{Esirkepov_1998,Farina01a,Farina05} and embedded in long laser
pulses \cite{Farina05,Sanchez_2011a,Sanchez_2011b} waves, were
found. Single-hump and multi-hump solitons are possible.
\Changes{All these solutions, except the one presented in Ref
\onlinecite{Esirkepov_1998} that admits an analytical form, were
found numerically}. The analysis of linearly polarized waves,
however, is more difficult mainly due to the high harmonic
generation. Some \Changes{approximate} one-dimensional solutions
have been presented in the weak amplitude limit
\cite{Hadsievski_02,Mancic_2006} and in the framework of the
Akhiezer-Polovin system \cite{Saxena_09}. Exact 2-dimensional
linearly polarized solitary wave solutions of the fluid model were
also found \cite{Sanchez_11}. Solitary waves were observed in
particle-in-cell (PIC) codes
\cite{Bulanov_92,Bulanov_99,Esirkepov_02} and their footprints were
detected in laboratory experiments
\cite{Borghesi_02,Chen_07,Pirozhkov_07,Sarri_10,Sylla_12}.

In the theoretical works, the analysis was mostly focused on the
spatio-temporal structures of the plasma and electromagnetic fields
and the relationship between the amplitude of the electromagnetic
wave and its frequency. A central role in the analysis was also
played by the organization of the solitons in parameter space, i.e.,
in the velocity and frequency plane. Some solitary waves  are
organized in branches: given a value of the group velocity, the
soliton exists for a single frequency value. Examples of this type
are the circularly polarized moving solitons discovered in Ref.
\onlinecite{Farina01a}. On the other hand, other types of solitons,
like the circularly polarized standing soliton found in Ref.
\onlinecite{Esirkepov_1998} or the embedded solitons of Ref.
\onlinecite{Sanchez_2011a} and \onlinecite{Sanchez_2011b}, have a
\emph{continuous spectrum}. In this case, for a given velocity,
solitons exist for any frequency within a certain range. The
discrete or continuous character of the spectrum can be anticipated
by using arguments from the dynamical system theory, including the
dimension of the system, the characteristics of the stable and
unstable manifolds of the fixed point where the orbit connects, and
the Hamiltonian and/or reversible character of the
system~\cite{Sanchez_2011a}.

A common feature to all the previously cited 1-dimensional solitons
is the stationary character of the soliton profile in a frame moving
with its group velocity. Under the traveling wave ansatz, the
amplitude of the solutions just depends on the coordinate
$\xi=x-Vt$. However, in recent fluid simulations aimed at the
interaction of an ultrashort laser pulse with an overdense plasma in
the relativistic transparency regime, solitonic structures with a
breather-like behavior were observed \cite{Wu_2013}. Energy exchange
between the soliton and the plasma occurred at twice the laser
frequency. These periodic oscillations in time of the soliton
amplitude is a novel aspect in one-dimensional fluid theory but a
well-known feature for linearly s-polarized 2-dimensional solitons,
as shown by past particle-in-cell simulation \cite{Bulanov_99} and
more recent solutions found in the fluid model \cite{Sanchez_2011a}.
An oscillatory  behavior of 1-dimensional linearly polarized
solitary waves was also observed in PIC simulations of laser-plasma
interactions \cite{Macchi2012}.

These observations in both fluid \cite{Wu_2013} and PIC simulations
\cite{Macchi2012} motivate \ESedit{us to determine numerically}
exact linearly polarized solitary wave solutions of the relativistic
fluid model. Following the findings of Ref. \cite{Wu_2013}, we did
not restrict the analysis to stationary solutions, but let them
oscillate periodically in time. From a numerical point of view, this
a challenging problem because one needs to work with a system of
partial differential equations (instead of ordinary differential
equations). \Changes{This method, unlike the excitation of solitary
waves with laser pulses in fluid or PIC simulations, provides the
\ESedit{frequency-amplitude relation} and the frequency range of
existence of the waves}. Section \ref{Sec:Fluid:Model} introduces
the relevant equations of the model and presents the numerical
method used to compute the solitary waves. Results on standing
solitary waves are presented in Sec. \ref{Sec:Soliton:Standing},
including their spatio-temporal structure, a simple analysis based
on energy conservation and the dependence of their main properties
as a function of the frequency. The stability is explored in Sec.
\ref{Sec:Stability} with the aid of full non-stationary fluid
simulations. The conclusions are summarized in Sec.
\ref{Sec:Conclusions}.

\section{Physical and numerical  model \label{Sec:Fluid:Model}}
\subsection{The relativistic fluid model}

We consider a plasma consisting of electrons and immobile ions. For
convenience, length, time, velocity, momentum, vector and scalar
potentials and density are normalized over $c/\omega_{pe}$,
$\omega_{pe}^{-1}$, $c$, $m_ec$, $m_ec^2/e$ and $n_0$, respectively.
Here $n_0$, $\omega_{pe}=\sqrt{4\pi n_0e^2/m_e}$, $m_e$ and $c$ are
the unperturbed plasma density, the electron plasma frequency, the
electron mass and the speed of light. Maxwell (in the Coulomb gauge)
and plasma equations then read
\begin{subequations}
\begin{equation}
\Delta \bm{A}-\frac{\partial^2\bm{A}}{\partial
t^2}-\frac{\partial}{\partial t}\nabla\phi = n\bm{v}\label{Eq:A}
\end{equation}
\begin{equation}
\Delta \phi = n-1\label{Eq:Poisson}
\end{equation}
\begin{equation}
\frac{\partial n}{\partial
t}+\nabla\cdot\left(n\bm{v}\right)=0\label{Eq:Continuity}
\end{equation}
\begin{equation}
\frac{\partial \bm{P}}{\partial
t}-\bm{v}\times\left(\nabla\times\bm{P}\right)=\nabla\left(\phi-\gamma\right)\label{Eq:P}
\end{equation}\label{Sys:Fluid:0}
\end{subequations}
where $\bm{A}$ and $\phi$ are the vector and scalar potentials,
$\bm{P}=\bm{p}-\bm{A}$, $\gamma=\sqrt{1+|\bm{p}|^2}$ and $\bm{p}$
and $\bm{v}=\bm{p}/\gamma$ are  the electron momentum and velocity,
respectively. \ESedit{In this model the plasma is assumed to be cold
since the typical background thermal velocities are much smaller
than the relativistic quiver velocity of electrons in the strong
electromagnetic field. Extensions of previous works on relativistic
solitons to include a finite temperature have shown that certain
feature of the solitons change, but that the cold fluid model
remains a useful first approximation
\cite{Poornakala_2002,Sanchez_2011a}.}

Here we restrict the analysis to  one dimensional waves
($\partial_y=\partial_z=0$) propagating along the $x$ direction.
Coulomb gauge and the transverse components of \refeq{Eq:P} give
$A_x=0$ and $P_y=P_z= 0$, respectively. Using these results in Sys.
\refeq{Sys:Fluid:0}, one finds
\begin{subequations}
\begin{equation}
\frac{\partial^2 \phi}{\partial t\partial
x}+\frac{n}{\gamma}p_x=0\label{Eq:Phi}
\end{equation}
\begin{equation}
\frac{\partial A_{y,z}}{\partial x^2}-\frac{\partial^2
A_{y,z}}{\partial t^2}- \frac{n}{\gamma}A_{y,z}=0\label{Eq:A:1}
\end{equation}
\begin{equation}
n = 1+\frac{\partial^2 \phi}{\partial x^2}
\end{equation}
\begin{equation}
\frac{\partial n}{\partial t}+\frac{\partial }{\partial
x}\left(\frac{np_x}{\gamma}\right)=0
\end{equation}
\begin{equation}
\frac{\partial p_x}{\partial t} = \frac{\partial}{\partial
x}\left(\phi-\gamma\right)\label{Eq:Phi0}
\end{equation}\label{Sys:Fluid:x}
\end{subequations}

\subsection{The Akhiezer-Polovin system\label{SubSec:Akhiezer}}

Before we discuss solitary waves solutions of Sys.
(\ref{Sys:Fluid:x}), we first review some concepts of the
Akhiezer-Polovin set of equations \cite{Akhiezer_56}. This system,
which is a subset of Sys. \refeq{Sys:Fluid:x}, is obtained by
assuming that all the variables depend on the coordinate $\xi =
x-Vt$, with $V\equiv \omega/k>1$ the normalized phase velocity. The
result is a set of two ordinary differential equations that are
simpler than Sys. (\ref{Sys:Fluid:x}). The purpose of this short
analysis is twofold. On one hand, analytical tools from the theory
of dynamical system show that linearly polarized and isolated
solitary waves depending only on $\xi$ are not possible. This
property suggests that one may look for solitary waves depending
separately on  $x$ and $t$. This procedure is followed in Sec.
\ref{Sec:Soliton:Standing}. On the other hand, the dispersion
relation of the  Akhiezer-Polovin system will help us to understand
the results of Sec. \ref{Sec:Soliton:Standing} \ESedit{concerning small
amplitude oscillations at the tail of the solitary wave.}

Taking linear polarization, $\bm{A} = a(\xi) \bm{u}_y$,
\refeq{Sys:Fluid:x} becomes
\begin{subequations}
\begin{equation}
\frac{d}{d\xi}\left[-V n+n \frac{p_x}{\gamma}\right]=0\ \ \ \
\ \rightarrow \ \ \ \ \ -V n+n \frac{p_x}{\gamma} =
-V\label{Eq:Aux1}
\end{equation}
\begin{equation}
\frac{d}{d\xi}\left[ V p_x+\phi-\gamma\right]=0 \ \ \ \ \
\rightarrow \ \ \ \ \ V p_x+\phi-\gamma = -\Gamma \label{Eq:Aux2}
\end{equation}
\begin{equation}
\frac{d^2 \phi}{d\xi^2}=n-1\label{Eq:Poisson:0}
\end{equation}

\begin{equation}
\left(V^2-1\right)\frac{d^2
a}{d\xi^2}+\frac{n}{\gamma}a=0\label{Eq:Wave:0}
\end{equation}
\end{subequations}
 where we assumed that at certain position, $\xi_0$,  one has
 \begin{equation}
 a(\xi_0) = a_0,\ \ \ \ \phi(\xi_0)=0, \ \ \ \ \ \
 p_x(\xi_0)=0,\ \ \ \ \ \
 n(\xi_0)=1\label{Eq:BC}
 \end{equation}
 and we introduced the constant $\Gamma\equiv \sqrt{1+a_0^2}$.
Defining $\psi\equiv\Gamma+\phi$ and $R \equiv
\sqrt{\psi^2-(1-V^2)(1+a^2)}$, equations \refeq{Eq:Aux1},
\refeq{Eq:Aux2} and $\gamma = \sqrt{1+a^2+p_x^2}$ give $p_x =
(V\psi-R)/(1-V^2)$, $n=V(\psi-VR)/R(1-V^2)$ and
$\gamma=(\psi-VR)/(1-V^2)$. The substitution of these results in
Eqs. \refeq{Eq:Poisson:0} and \refeq{Eq:Wave:0} yields
\cite{Akhiezer_56,Lehmann_11}
\begin{subequations}
\begin{equation}
\frac{d^2a}{d\xi^2}=-\frac{V}{V^2-1}\frac{a}{R}
\end{equation}
\begin{equation}
\frac{d^2\phi}{d\xi^2}=-\frac{V}{V^2-1}\left(\frac{\psi}{R}-\frac{1}{V}\right)
\end{equation}\label{Sys:Akhiezer}
\end{subequations}
The above system can be written in Hamiltonian form. The Hamiltonian
function
\begin{equation}
H =
\frac{1}{2}\left(V^2-1\right)\dot{a}^2+\frac{1}{2}\dot{\phi}^2+\frac{V}{V^2-1}\left(R-V\right)-\frac{\phi}{V^2-1}\label{Eq:Hamiltonian},
\end{equation}
which does not depend explicitly on $\xi$, is a constant of motion.
In the small but finite amplitude limit with low density plasma
($V\rightarrow 1$), \refeq{Sys:Akhiezer} shows that the dispersion
relation (in physical units) for the transverse oscillations is
given by\cite{Akhiezer_56}
\begin{equation}
\omega\approx
kc+\frac{\omega_{pe}^2}{2kc}\left(1-\frac{a_0^2}{2}\right)\label{Eq:Dispersion}
\end{equation}
Higher order corrections can be found in Ref. \cite{Pesch_07}. The
longitudinal variable, $\phi$, oscillates with frequency $2\omega$
and an amplitude of the order of $a_0^2$.

System \refeq{Sys:Akhiezer} can be written as
$d\bm{y}/d\xi=\bm{f}(\bm{y})$ and arguments from the theory of
dynamical systems can be used to discuss the existence of solitary
waves \cite{Champneys97}. We introduce the notion of the stable
(unstable) manifold $W^s$ ($W^i$ ) of an invariant set
$\mathcal{M}$, as the set of forward (backward) in $\xi$
trajectories that terminate at $\mathcal{M}$. The invariant set of
interest here are the equilibrium points $\bm{y}^*$ and the periodic
orbits $\bm{y}_p$, that satisfy $\bm{f}(\bm{y}^*)=0$, and
$\bm{y}_p(t)=\bm{y}_p(t+T)$, respectively. Since solitary waves are
homoclinic orbits, i.e. they approach $\mathcal{M}$ as
$\xi\rightarrow \pm \infty$, they lie in the intersection of $W^s$
and $W^i$. A local analysis of the stability of $\bm{y}^*$ or
$\bm{y}_p$ can help to decide the existence of solitary waves. For
example, if there is no $W^s$ and $W^i$ because the equilibrium
point is a center, then homoclinic orbits cannot exist.

If $a_0\ne 0$, then the phase space point
$(a,\dot{a},\phi,\dot{\phi})=(a_0,0,0,0)\equiv Q_0$ is not an
equilibrium state of  system \refeq{Sys:Akhiezer}; solitary waves
with $a\rightarrow a_0$ as $\xi\rightarrow  \pm \infty $ are not
possible. On the other hand, if $a_0=0$, then $Q_0$ is an
equilibrium state and one may try to look for solitary waves, i.e.
homoclinic orbits connecting to $Q_0$ at $\xi \rightarrow \pm
\infty$. However, orbits cannot connect to $Q_0$ because a standard
stability analysis of $Q_0$ shows that it is a center with
eigenvalues $\lambda_{1,2}=\pm i\sqrt{1/(V^2-1)}$ and
$\lambda_{3,4}=\pm i/V$. By going back to physical units, one
verifies that ``frequency'' $1/V$ corresponds to longitudinal
oscillations at the plasma frequency $\omega_{pe}$. The
``frequency'' $\sqrt{1/(V^2-1)}$ corresponds to linear
electromagnetic waves and it reduces to \refeq{Eq:Dispersion} in
the small amplitude and low plasma density limit.

The above discussion rules out the existence of homoclinic orbits to
$Q_0$. From a physical point of view they would represent isolated
solitary waves, with linear polarization, and stationary in a frame
of reference moving with $V$. However, as previously discussed, one
could also try to look for homoclinic orbits connecting to a
periodic solution as $\xi \rightarrow \pm \infty$. These structures,
that were constructed using Poincar\'e analysis in Ref.
\cite{Saxena_09}, are solitary waves embedded in long laser pulses.
We remark that, since we used \refeq{Eq:BC} to derive
\refeq{Sys:Akhiezer}, only orbits of Sys. (\ref{Sys:Akhiezer})
passing through the surface
$(a,\dot{a},\phi,\dot{\phi})=(a_0,\dot{a},0,\dot{\phi})$ are
relevant from a physical point of view. Other orbits would be a
solution of \refeq{Sys:Akhiezer} but not of the fluid model
\refeq{Sys:Fluid:0}.

\subsection{Locating generalized solitary waves\label{SubSec:Numerical}}

A new class of solitary waves is found if we let the solutions vary
with  time in the $\xi$-frame. Physically, such a broader model is
necessary to take into account the appearance of new frequencies in
the solution. It requires a more general mathematical framework than
the Akhiezer-Polovin system. For this purpose, scaled spatial and
temporal variables $X = x/L$ and $\tau=t/T$ are introduced, with $T$
and $L$ the temporal and spatial periods of the solution. System
(\ref{Sys:Fluid:x}) then reads
\begin{subequations}
\begin{equation}
\frac{1}{L^2}\frac{\partial^2 A_{y,z}}{\partial
X^2}-\frac{1}{T^2}\frac{\partial^2 A_{y,z}}{\partial \tau^2}-
\frac{n}{\gamma}A_{y,z}=0\label{Eq:A:2}
\end{equation}
\begin{equation}
\frac{1}{T^2}\frac{\partial^2 p_x}{\partial
\tau^2}+\frac{1}{TL}\frac{\partial^2\gamma}{\partial X\partial
\tau}+\frac{n}{\gamma}p_x = 0\label{Eq:Px:2}
\end{equation}\label{Sys:Numerical}
\end{subequations}
 \begin{subequations}
 \begin{equation}
n = 1+\frac{1}{TL}\frac{\partial^2 p_x}{\partial X\partial
\tau}+\frac{1}{L^2}\frac{\partial^2\gamma}{\partial X^2}
 \end{equation}
where
\begin{equation}
\gamma = \sqrt{1+p_x^2+A_y^2+A_z^2}
\end{equation}
\end{subequations}
and \refeq{Eq:Phi0} was used to eliminate $\phi$. We remark that $T$
and $L$ appear now as free parameters in \refeq{Sys:Numerical}. For
convenience, the solitary wave frequency $\Omega=2\pi/T$ (instead of
$T$) will be used as a bifurcation parameter to present our results.
As explained below, $L$ is found with the aid of a phase condition
for each $\Omega$ value.

Our algorithm solves \refeq{Sys:Numerical} as follows. The
computational box  $(0<X<1)\times (0<\tau<1)$ is discretized with
$N_x\times N_\tau$ uniformly distributed points. Here $N_x$ and
$N_\tau$ are odd numbers. A state vector $\bm{x}_s=[A_y\ A_z\ p_x\
L]$ of dimension $3\times N_x\times N_\tau+1$ is constructed. It
contains the unknowns of the problem, i.e., the values of $A_y$,
$A_z$, and $p_x$ at the grid points and the length of the
computational domain. Differential operators in
\refeq{Sys:Numerical} are substituted by second order finite
difference formulas. At the border of the box one may take into
account the  boundary conditions of the problem. Here we are
interested in periodic in time solutions
$A_{y,z}(X,\tau+T)=A_{y,z}(X,\tau)$ and
$p_{x}(X,\tau+T)=p_{x}(X,\tau)$ and two types of spatial boundary
conditions: (i) vanishing
$A_{y,z}(0,\tau)=A_{y,z}(L,\tau)=p_x(0,\tau)=p_x(L,\tau)=0$ and (ii)
periodic $A_{y,z}(X,\tau)=A_{y,z}(X+L,\tau)$ and
$p_{x}(X,\tau)=p_{x}(X+L,\tau)$. The use of finite differences
transforms \refeq{Sys:Numerical} into a system of $3N_x\times
N_\tau$ nonlinear algebraic equations.

The algorithm is closed by taking into account that the  problem is
invariant under temporal and spatial translations. Its solution is
not unique, but rather is a continuous family. In order to remove
this arbitrary \emph{phase variable} and restrict the solution to a unique
member of the family, a \emph{phase condition} was added. We found
that
\begin{equation}
\frac{\partial^2 A_y }{\partial X\partial \tau}\mid_{X =
0.5,\tau=0.5} =0\label{Eq:Phase}
\end{equation}
\ESedit{serves} this purpose \Changes{and allows the code to
converge to the solution in few iterations}. Equation
\refeq{Eq:Phase} was also approximated by the corresponding
finite-difference formula.

The set of $3N_x\times N_\tau+1$ algebraic equations,
\begin{equation}
\bm{F}(\bm{x}_s)=0\label{Eq:Newton},
\end{equation}
was solved with a Newton-Raphson method. The speed of the algorithm
was enhanced by computing analytically the Jacobian, $\bar{\bm{J}}$,
of $\bm{F}(\bm{x}_s)$ and using parallel computations to find the LU
decomposition of $\bar{\bm{J}}$ for each iteration. The tolerance of
our solutions, $Error = max\left[\bm{F}(\bm{x}_s)\right]$, was less
than $10^{-8}$. The initial guess used to initialize the Newton-Raphson method
depends on the solution under consideration, as discussed shortly.

\begin{figure}[h]
\begin{center}
\includegraphics[scale=0.62]{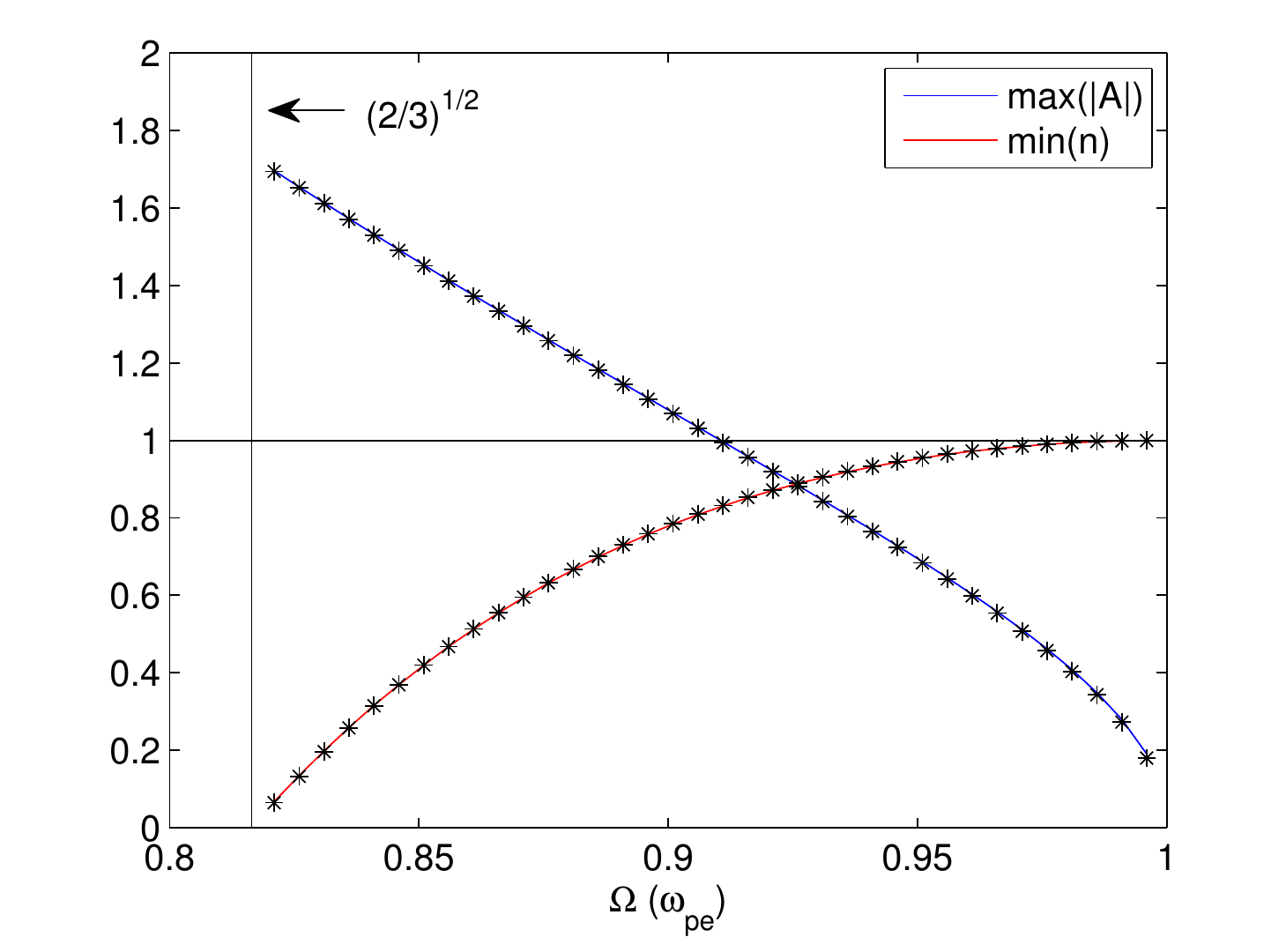}
\end{center}
\caption{\label{Fig:Comparison:1}\Changes{(color online)} Maximum
amplitude, $max(|\bm{A}|)$, and minimum density, $min(n)$, for
circularly polarized solitary waves. Circles (solid lines)
correspond to solutions computed with our numerical code (analytical
solution given in Ref. \cite{Esirkepov_1998})}
\end{figure}

The code was validated with the circularly polarized solitary waves
obtained by Esirkepov et al \cite{Esirkepov_1998}. For these
calculations we set  $N_x = 1000$ and $N_\tau=60$. The comparison
between the analytical formula given in Ref. \cite{Esirkepov_1998}
and the result of our code is shown in Fig. \ref{Fig:Comparison:1}.
At each $\Omega$, we used as initial guess for the Newton method the
solution obtained at the previous $\Omega$ value in the branch. For
the first frequency, $\Omega=0.996$, the analytical formula was
used. The calculations were carried out with periodic and vanishing
boundary conditions in space. Both schemes yielded the same results,
that are in good agreement with the analytical formulae. For this
solution one has $|\bm{A}|\rightarrow 0$ as $\Omega\rightarrow
\omega_{pe}$ and the branch extends until
$\Omega=\sqrt{2/3}\omega_{pe}$, where the minimum density vanishes.

\section{Breather-like solitary waves \label{Sec:Soliton:Standing}}

\subsection{Branch of solutions\label{SubSec:Branch}}

We now focus on linearly polarized solitary waves and set $A_z=0$ in
Eqs. \refeq{Eq:A:2} and \refeq{Eq:Px:2}. The number of unknowns in
Sys. \refeq{Eq:Newton} is equal to $2N_xN_\tau+1$, thus reducing the
computational load as compared with the circularly polarized case.
The guess to initialize the Newton-Raphson algorithm was taken from Ref.
\cite{Mancic_2006} that presented an analytical formula for linearly
polarized solitary waves. This solution is not exact but only valid
for frequency close to $\omega_{pe}$, where the amplitude of the
wave vanishes. In this limit, the authors derived a nonlinear
Schr\"{o}dinger equation with local and non-local cubic
nonlinearities and found a standing electromagnetic solitary wave.
Fortunately, this solution is \emph{close} enough to our
breather-like wave. Here \emph{close} means that our Newton method
started with this wave converges to a solution during the iterative
process. Once a breather-like solution is known for a $\Omega$ value
close to $\omega_{pe}$, the branch of solution is continued by using
$\Omega$ as a bifurcation parameter. For each parameter value the
spatio-temporal structure of the solitary wave was obtained. Typical
resolutions in the calculations were $N_x = 1001$ and $N_\tau = 101$
and the length of the computational box dynamically changed during
the calculation within the range $100<L<200$. To check the integrity
of the solutions, the same calculations but with different
resolutions were also carried out.

\begin{figure}[h]
\begin{center}
\includegraphics[scale=0.62]{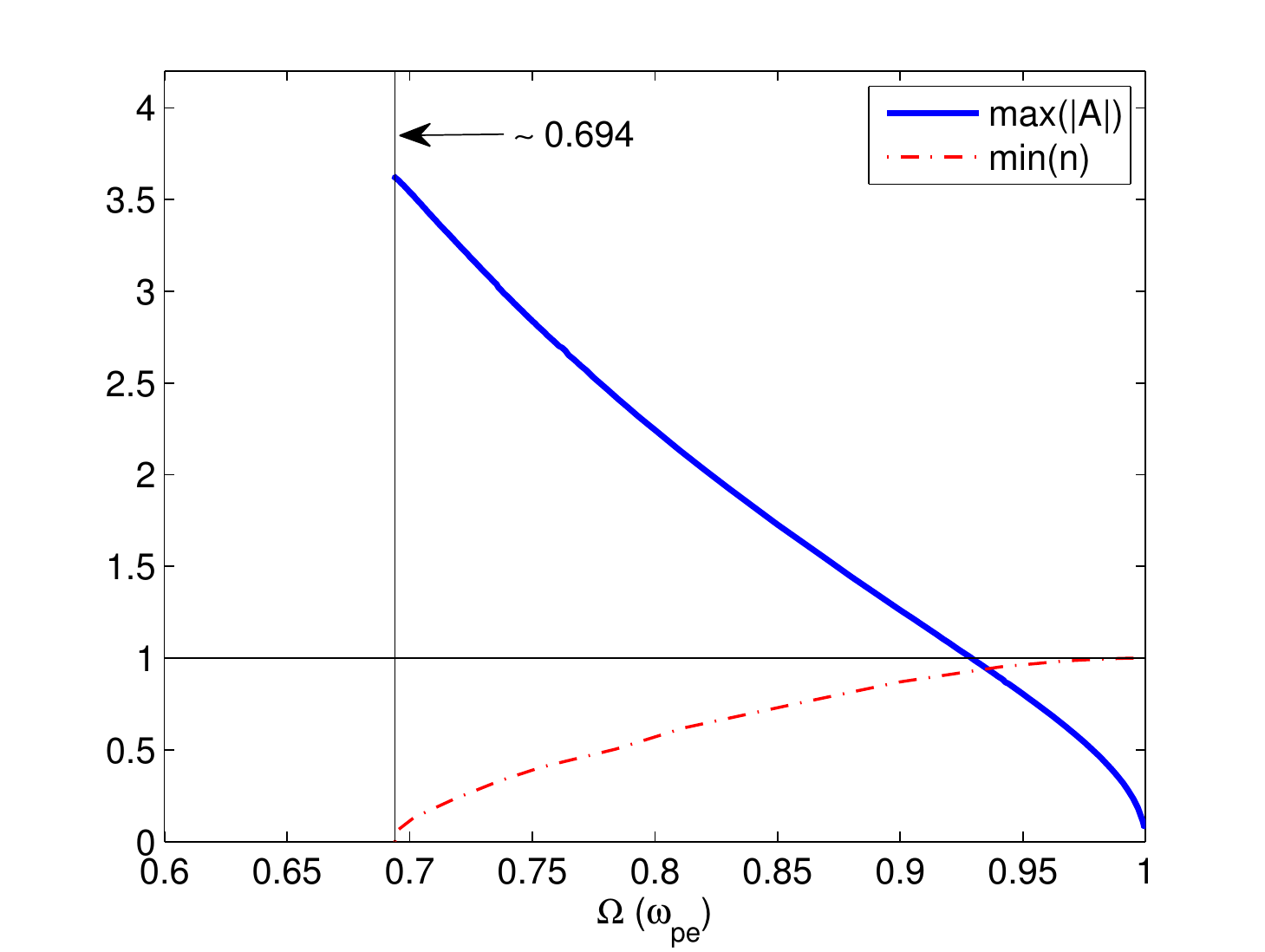}
\end{center}
\caption{\label{Fig:Standing:Omega}\Changes{(color online)} Maximum
amplitude (solid blue line ) and minimum density (dashed red line)
values versus the normalized frequency for linearly polarized
solitary waves.}
\end{figure}

Figure \ref{Fig:Standing:Omega} shows the maximum of the vector
potential (blue thick line ) and the minimum density (dashed red
line) values of the solitary waves versus the frequency $\Omega$.
The behavior is qualitatively similar to the circularly polarized
case (see Fig. \ref{Fig:Comparison:1}): (i) solitary waves exist in
a frequency range $\Omega_{min}<\Omega<\omega_{pe}$, (ii)
$max(|A|)\rightarrow 0$, $min(n)\rightarrow 1$, and the width
increases as $\Omega\rightarrow \omega_{pe}$, and (iii) the maximum
of the amplitude increases and $min(n)\rightarrow 0$ as the
frequency approaches $\Omega_{min}\approx 0.694$.
A comparison of Fig.~\ref{Fig:Comparison:1} and \ref{Fig:Standing:Omega} shows that, for
a given frequency, linearly polarized waves have greater amplitude
than  circularly polarized waves. They also exist in a broader
frequency range because $\Omega_{min}=0.694 < \sqrt{2/3}\approx
0.817$.

\subsection{Spatio-temporal structure of the waves \label{SubSec:Structure}}

\begin{figure*}
\includegraphics[scale=0.8]{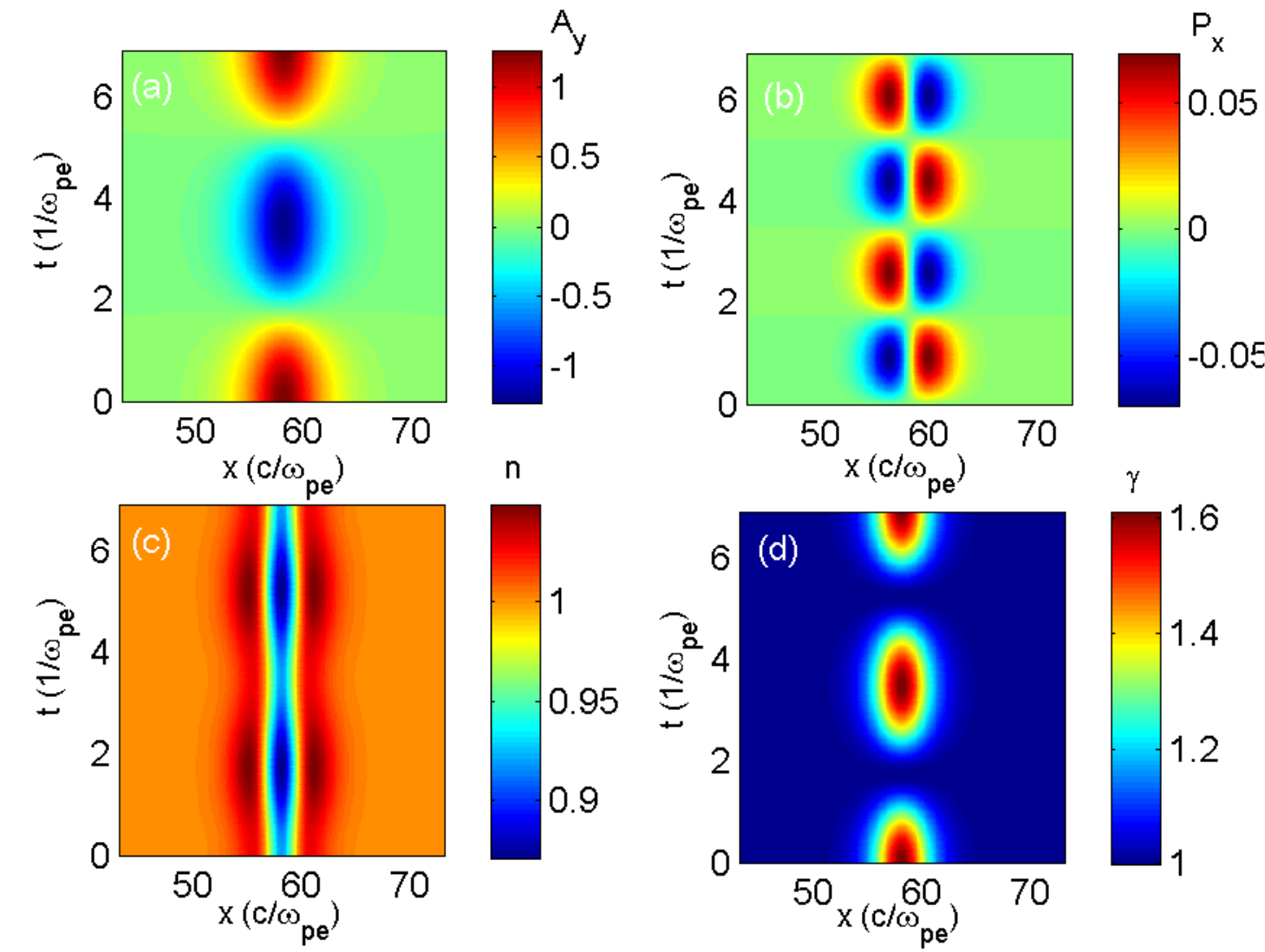}
\caption{\label{Fig:Sol:09}\Changes{(color online)} Spatio-temporal
evolution of a linearly polarized soliton with
$\Omega=0.9\omega_{pe}$. Numerical values are $N_x = 1001$, $N_\tau
= 101$. Panels (a) to (d) show the vector potential component $A_y$,
the longitudinal momentum $p_x$, the density $n$ and $\gamma$,
respectively. }
\end{figure*}

\begin{figure*}
\includegraphics[scale=0.8]{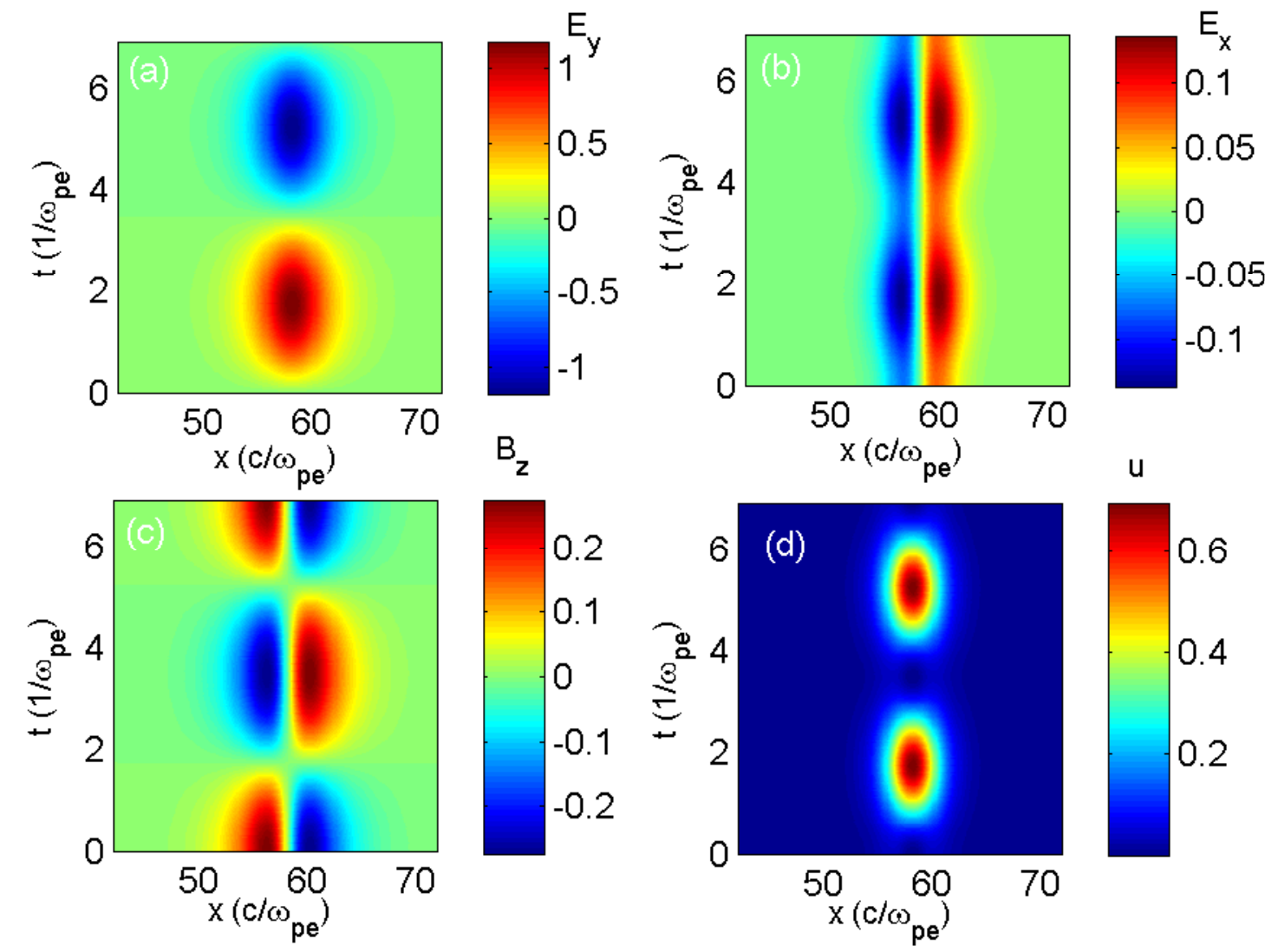}
\caption{\label{Fig:Sol:09:Fields}\Changes{(color online)
Panels (a) to (d) show the
(normalized) electric field components $E_y$, $E_x$, the magnetic
field component $b_z$ and the electromagnetic energy density $u =
(E_x^2+E_y^2+B_z^2)/2$, respectively, for the same parameter values
as in Fig.~\ref{Fig:Sol:09}.} }
\end{figure*}

Figure \ref{Fig:Sol:09} shows the spatio-temporal evolution of a
solitary wave with $\Omega=0.9\omega_{pe}$. Similarly to circularly
polarized solitons, there is a rarified plasma region and an
electromagnetic wave oscillating inside [see panels (a) and (c)].
However, linearly polarized waves present two distinguishing
features: (i) the plasma cavity has a time modulation [see panel
(c)] and (ii) longitudinal electron momentum $p_x$, electron density
and $\gamma$ factor oscillates twice faster than the vector
potential. From \refeq{Eq:Phi0}, we observe that electron
oscillations come from the actions of the electrostatic force due to
the charge separation and the ponderomotive force. The momentum in
panel (b) oscillates twice faster because of the second-harmonic
oscillating component of the ponderomotive force. The modulation in
time of the plasma cavity is correlated with the behavior of the
vector potential: as shown by panels (a) and (c), the lowest plasma
density inside the cavity is reached when the electromagnetic wave
vanishes.

\begin{figure}
\includegraphics[scale=0.3]{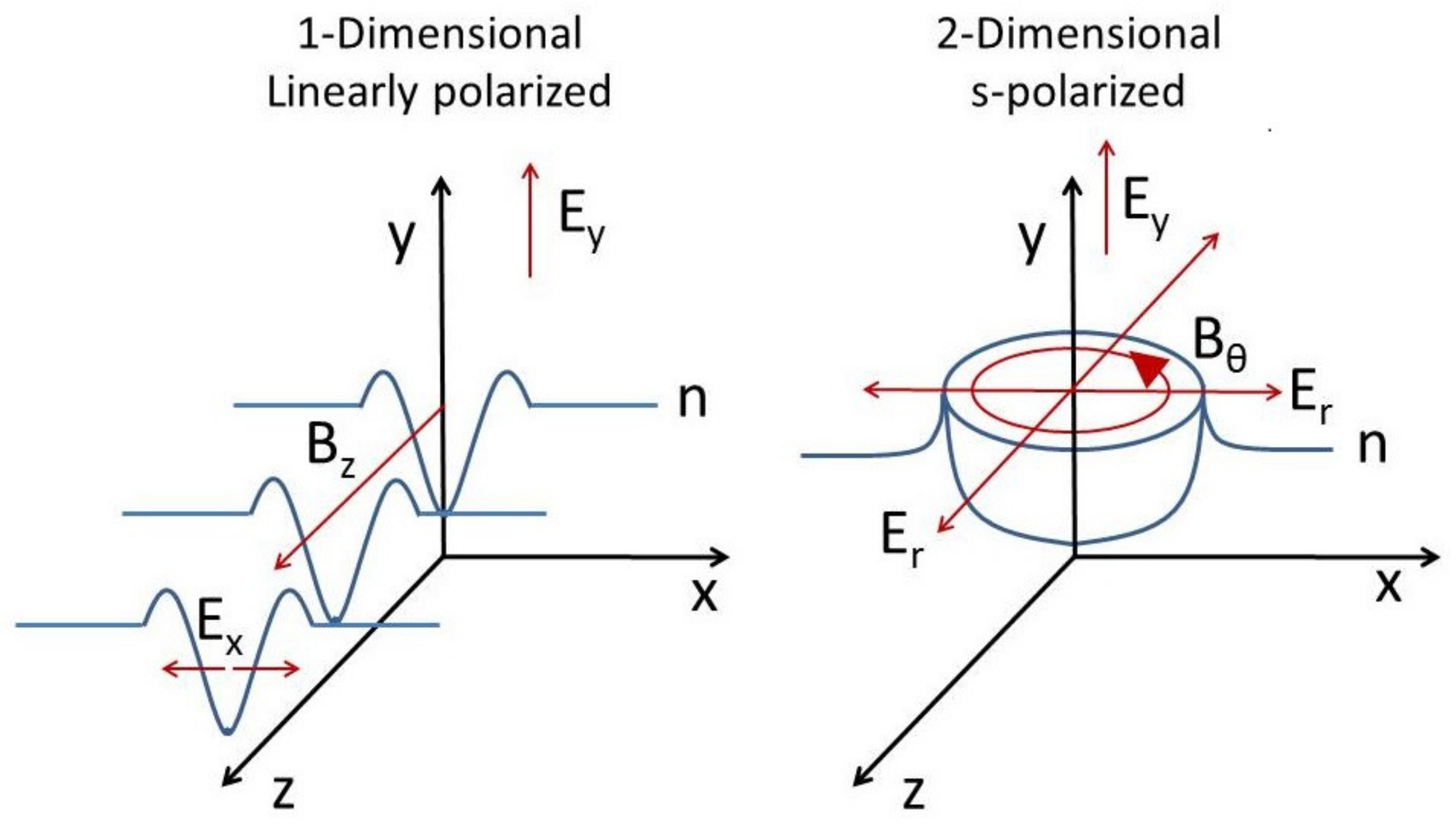}
\caption{\label{Fig:Waves1d2d}\Changes{(color online) Schematics of
the electron plasma density and the electromagnetic fields of 1 and
2-dimensional solitary waves with linear polarization
$\bm{A}=A_y\bm{j}$.} }
\end{figure}

\Changes{An analysis of the electromagnetic fields and the
electromagnetic energy density ($u=(|\bm{E}|^2+|\bm{B}|^2)/2$) helps
to understand the physics of the solitary waves and their connection
or generalization to the s-polarized two-dimensional solitary waves
studied in Refs. \onlinecite{Bulanov_99} and \onlinecite{Sanchez_11}
(see Figs \ref{Fig:Sol:09:Fields} and \ref{Fig:Waves1d2d}). In
either 1-dimensional (1D) or 2-dimensional (2D) geometry, the vector
potential and electric field components $A_y$ and $E_y$ oscillate at
the frequency of the solitary wave (below the plasma frequency). The
electromagnetic wave is trapped inside the cavity. They have
electric field components, $E_x$ for 1D and a radial component $E_r$
for 2D waves, that always point outside of the solitary wave. These
electric field components produce a force that tries to bring the
electrons back to the center of the wave and, unlike $E_y$, their
time variation is very weak [see panel (b) of Fig.
\ref{Fig:Sol:09:Fields}]. Since $\bm{p}_y=A_y$, there is a current
in the $y$-direction that produce a $B_z$ component and an azimuthal
component $B_\theta$ for 1D and 2D solitary waves, respectively. }

\Changes{As shown by Fig. \ref{Fig:Waves1d2d} one of the main
consequences of removing the hypothesis $\partial/\partial z =0$,
and thus  changing from a 1D to a 2D configuration, is the
confinement of the magnetic field. For 1D waves $B_z$ extends from
$z\rightarrow -\infty$ to  $z\rightarrow +\infty$ but for 2D it
becomes azimuthal and remains inside the cavity. In a 3D
configuration, where the electric field component $E_y$ cannot
extend  from $y\rightarrow -\infty$ to  $y\rightarrow +\infty$, a
change of topology happens in order to confine the full structure
inside a \ESedit{bubble}. Such a structure has been observed in PIC
simulations \cite{Esirkepov_02}. However,  to the best of our
knowledge, they have not been investigated by looking for exact
solutions of the fluid model (as it was done in this work for 1D and
in Ref. \onlinecite{Sanchez_11} for 2D waves). }

\begin{figure}[h]
\begin{center}
\includegraphics[scale=0.62]{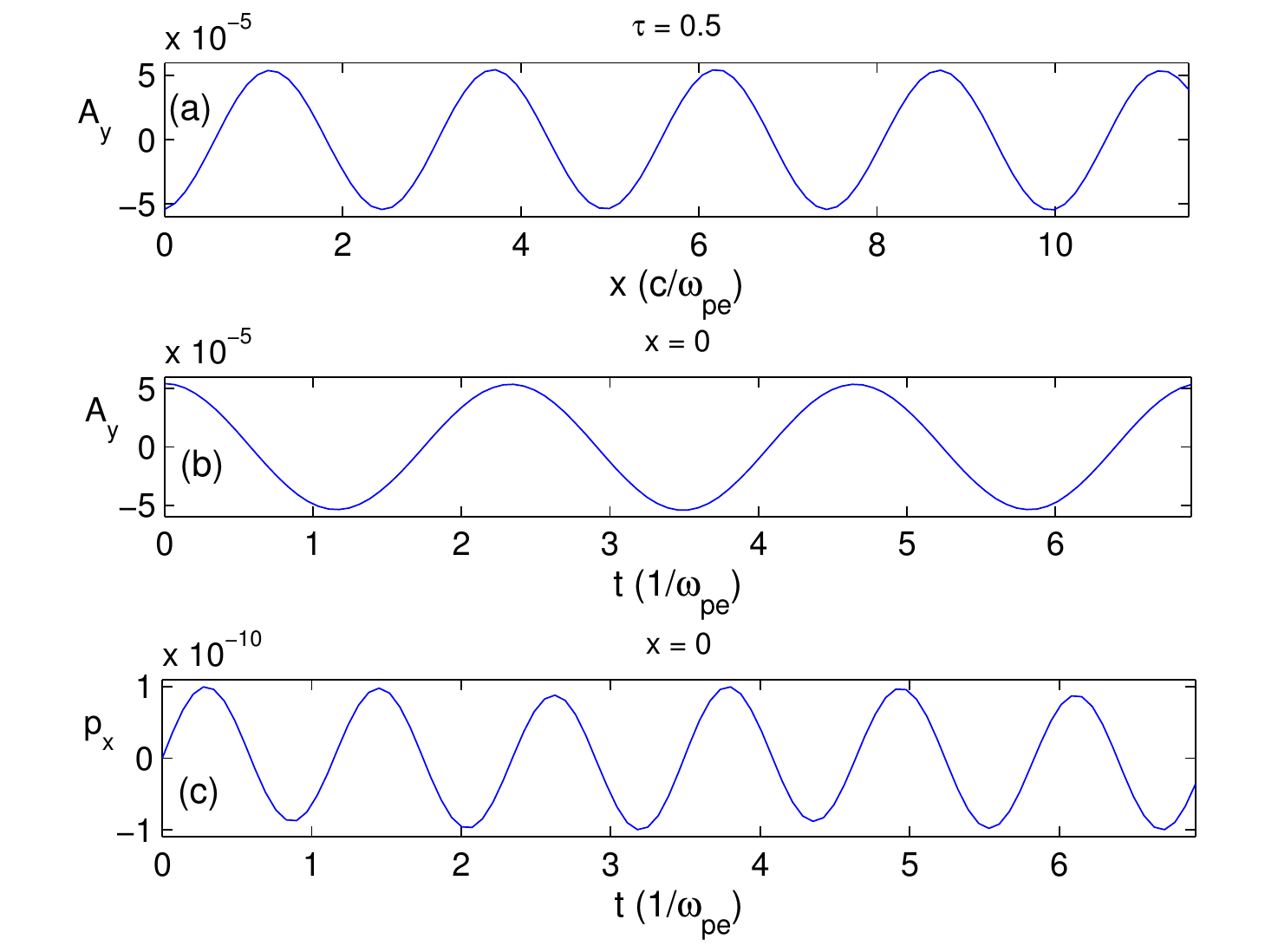}
\end{center}
\caption{\label{Fig:Soliton09:Tail}\Changes{(color online)} Vector
potential at the tail of a solitary wave with
$\Omega=0.9\omega_{pe}$. Panel (a) shows the spatial structure at
$\tau=0.5$ and panel (b) the temporal behavior at $x = 0$. }
\end{figure}

\ESedit{The maps of Fig. \ref{Fig:Sol:09} suggest the existence of
localized, time-periodic structures for which all plasma and
electromagnetic perturbations vanish as $X\rightarrow \pm \infty$).
However, we note that in our numerical solutions small amplitude,
periodic (standing wave) oscillations of the plasma and field
variables exist at the tails of a linearly polarized solitary wave.}
This is shown in \Changes{Fig. \ref{Fig:Soliton09:Tail}}, which
displays the spatial structure of the tail of the wave at certain
instant \ESedit{and the temporal structure at certain position for
$\Omega = 0.9\omega_{pe}$.} Low amplitude oscillations, \ESedit{of
order $5\times 10^{-5}$}, in the vector potential are evident in the
solitary wave's tails. \ESedit{These small amplitude oscillations
are connected to the periodic boundary conditions in our numerical
formulation, since they can be thought as corresponding to an
interference pattern between outgoing waves which re-enter the
computational box from both sides.}

\begin{figure*}
\includegraphics[scale=0.8]{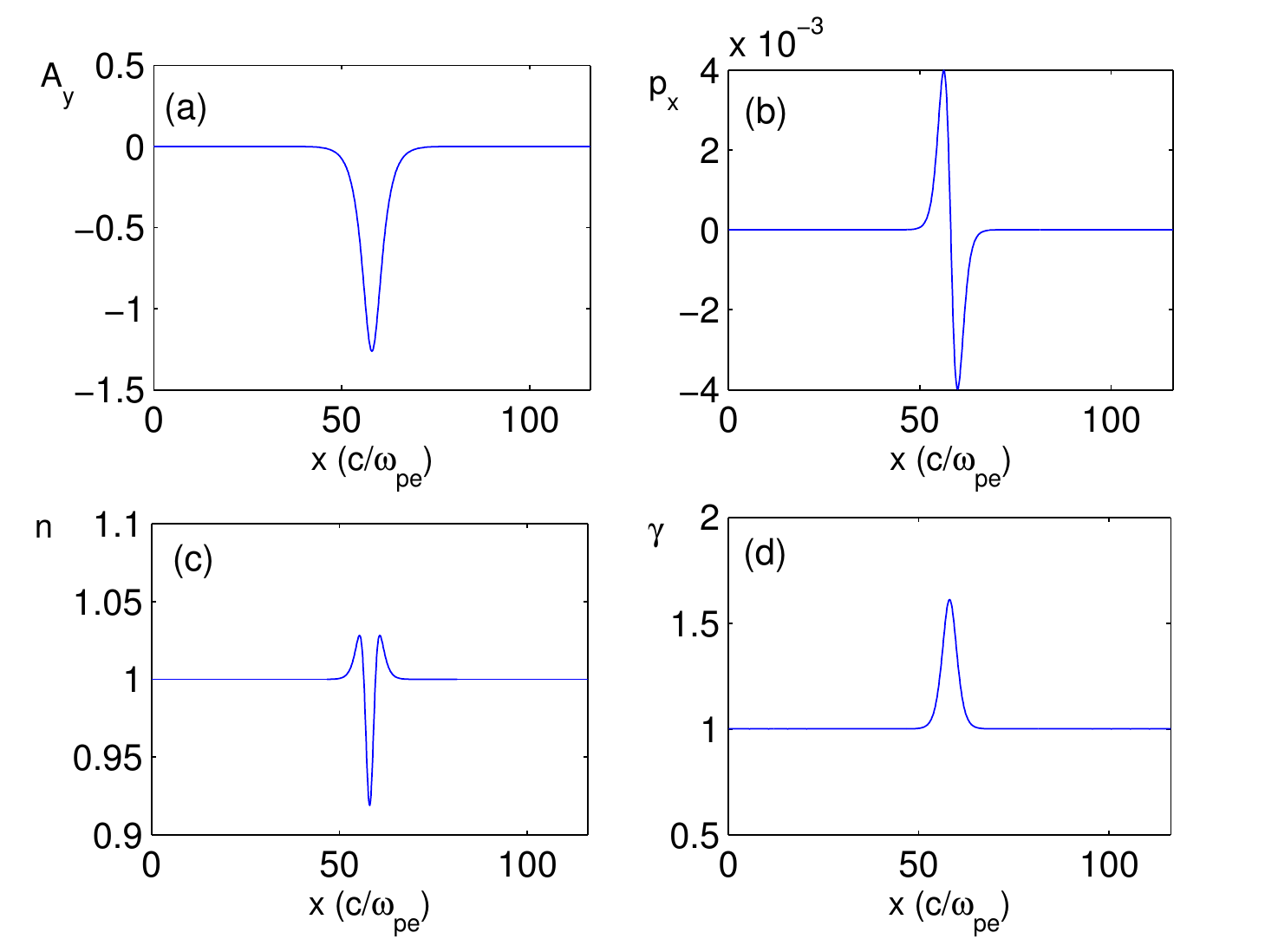}
\caption{\label{Fig:Sol:09:Plot}\Changes{(color online)} Spatial
structure of a linearly polarized soliton with
$\Omega=0.9\omega_{pe}$ at certain instant. Panels (a) to (d) show
$A_y$, $p_x$, $n$ and $\gamma$, respectively.}
\end{figure*}

\ESedit{We have found that, at a given frequency,
the precise amplitude of the small oscillations
depends on the initial guess, domain length and resolution of our
Newton method, while at the same time the localized solitary wave
structure remained the same. In particular, we have found oscillation
amplitudes in the range $10^{-5}\--10^{-2}$, with no indication for a dependence
of the amplitude on $\Omega$. This indicates that, while the amplitude and
profile of (the localized part of) the solitary wave is uniquely determined by
$\Omega$, there exists a continuum of solutions corresponding to tail
oscillations of different amplitude.}

\ESedit{Nevertheless, the standing wave oscillations in the tail are not completely
arbitrary, but actually obey the dispersion relation of the
Akhiezer-Polovin system, which is a particular case of Sys.~\ref{Sys:Numerical}. }
In the soliton's tail  $A_y$ and $\phi$ are
very small, thus indicating that the dynamics occurs close to the
Akhiezer-Polovin equilibrium point  $Q_0\equiv
(a,\dot{a},\phi,\dot{\phi})=(0,0,0,0)$. \ESedit{Around this equilibrium point
oscillations following the dispersion relation given by \refeq{Eq:Dispersion} may
occur. This statement is confirmed by looking closer to the small
oscillations in the tail of a wave for the particular case of Fig.~\ref{Fig:Soliton09:Tail}}.
From
panels (a) and (b), which show the spatial structure at $\tau=0.5$
and the temporal behavior at $x = 0$ of the vector potential, we
find the frequency $\omega\approx 2.67\omega_{pe}$, the wavenumber
$k\approx 2.51 c/\omega_{pe}$, and $a_0\approx 5\times 10^{-5}$.
Using $k= 2.51 c/\omega_{pe}$ in \refeq{Eq:Dispersion} yields $\omega\approx 2.71
\omega_{pe}$, which is in good agreement with $\omega\approx 2.67
\omega_{pe}$. As expected, the longitudinal variable $p_x$ has an
amplitude of the order of $\sim a_0^2$ and oscillates with frequency
$2\,\omega$ [see panel (c)]. \ESedit{This argument also indicates why the amplitude
of the tail oscillations is not uniquely determined: standing wave solutions
satisfying \refeq{Eq:Dispersion} may be found for any (sufficiently small) $a_0$.
For this reason, we consider the solitary waves presented here as 
localized structures.
}

\subsection{Energy balance\label{SubSec:Energy}}

Linearly polarized solitary waves are localized structures where
electromagnetic energy and electron kinetic energy are exchanged
periodically. This is an important difference compared to the
circularly polarized waves, that exhibit a stationary character.
Energy evolution is here analyzed using Sys. \refeq{Sys:Fluid:x}. It
conserves the sum of the three normalized energies, $E=E_l+E_p+E_e$,
where
\begin{align}
E_l =& \frac{1}{2}\int \left[\left(\frac{\partial A_y}{\partial
t}\right)^2+\left(\frac{\partial A_y}{\partial
x}\right)^2\right]dx\\
E_p = &\frac{1}{2}\int \left(\frac{\partial \phi}{\partial
x}\right)^2dx\\
 E_e =& \int \left(\gamma-1\right)n\ dx.
\end{align}
Here $E_l$, $E_p$ and $E_e$ are the energy of the electromagnetic
wave, the energy of the electrostatic plasma wave and the kinetic
energy of the electrons, respectively.

\begin{figure}[h]
\begin{center}
\includegraphics[scale=0.62]{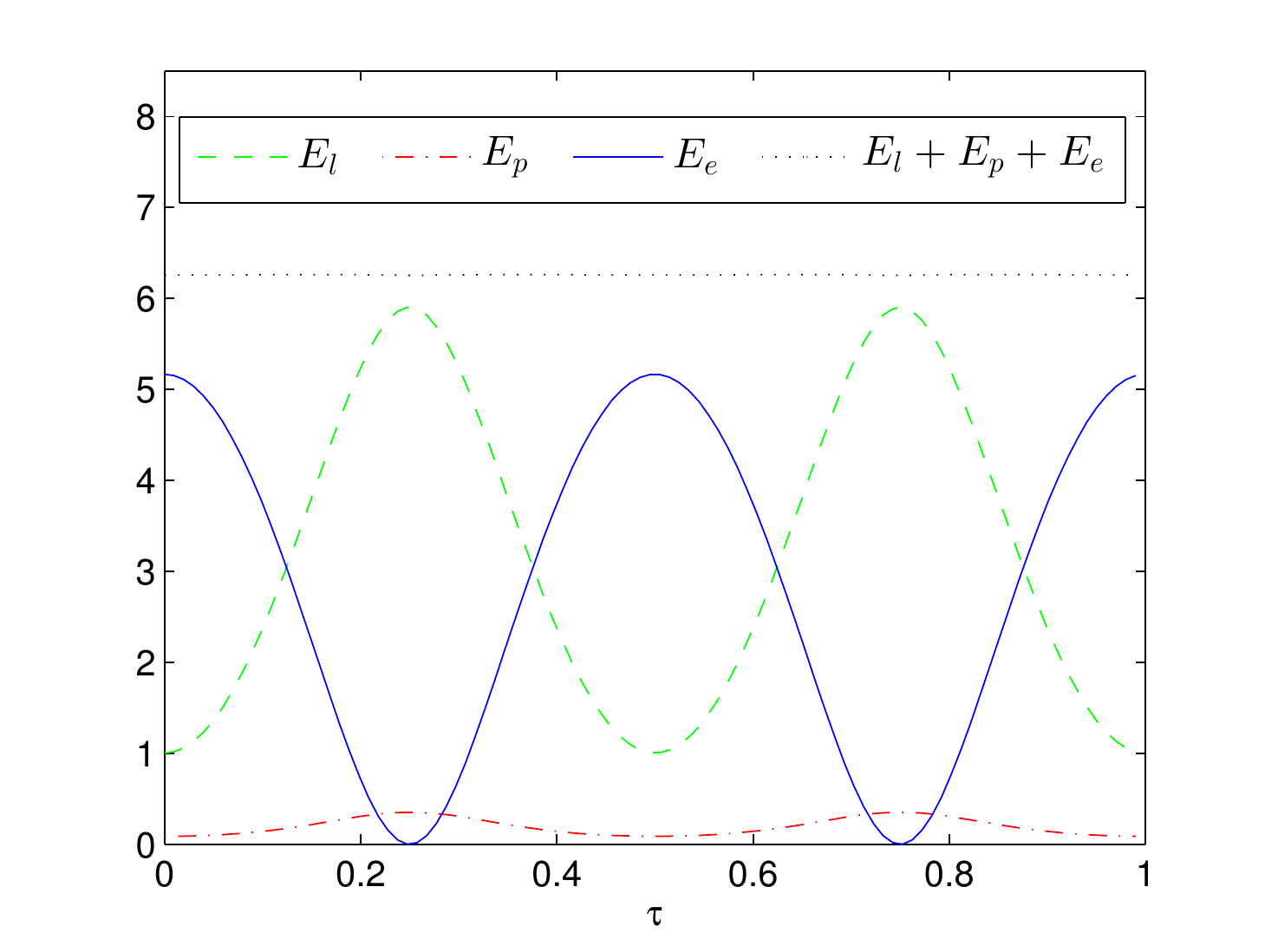}
\end{center}
\caption{\label{Fig:Soliton08:Energy}\Changes{(color online)}
Evolution of the energy of the electromagnetic wave ($E_l$), the
energy of the electrostatic plasma wave ($E_p$) and the kinetic
energy of the electrons ($E_e$) for a solitary wave with frequency
$\Omega=0.8\omega_{pe}$. The black line represents the sum of the
three terms.}
\end{figure}

As observed in Ref. \cite{Wu_2013}, linearly polarized breather-like
solitary waves exhibit a periodic exchange of energy. This feature
is shown in Fig. \ref{Fig:Soliton08:Energy}, which displays the time
evolution of the three energy contributions for a solitary wave with
$\Omega = 0.8\omega_{pe}$. Each period, the energy exchange is
repeated twice; the two minimums in the electron kinetic energy
coincide with the two maxima in the electromagnetic and plasma wave
energies. The electrostatic energy, albeit being the smallest
contribution, never vanishes, thus helping to maintain the plasma
cavity.

\section{Direct numerical simulation of the solitary waves \label{Sec:Stability}}

Having discussed a new family of breather like solitary wave
solutions in the previous section, we now turn to direct
fluid-Maxwell simulations of solitary waves of this family, in order
to validate the numerical procedure that we used to detect these
solutions. Moreover, since we integrate the fluid-Maxwell equations
for several periods of the solitary wave, we expect that if these
waves are prone to instabilities, these will be triggered within the
simulation time. A more rigorous stability study is beyond the scope
of this work.
The simulations involve numerically solving the full relativistic
fluid-Maxwell model [Sys.  (\ref{Sys:Fluid:x})]. We have used
two different codes for this purpose, a finite difference code and
a pseudospectral one.

As all the time
evolution equations we wish to solve here are either in continuity
or convective form, it was very convenient to employ the subroutine
package LCPFCT for the finite difference solver.
LCPFCT is a freely available
subroutine package developed at Naval Research Laboratory (NRL),
USA~\cite{lcpfct}. These subroutines are based on the principle of flux-corrected
transport \cite{boris}. Periodic boundary conditions are implemented
for the simulations presented here and Courant stability condition
is used to calculate appropriate value of integration time step for
ensuring the numerical stability. Profiles of fluid variables are
specified at the grid centers whereas the interface values of flow
variables is used. The flux-correction method has been quite
successful in solving fluid flow problems and it ensures density
positivity as well as numerical accuracy.

The pseudospectral code has been presented
in Ref.~\cite{siminos2014} and uses Fourier space discretization
of the partial derivative with respect to space and an adaptive,
fourth order Runge-Kutta scheme for time stepping.

\begin{figure*}
\begin{center}
\includegraphics[scale=0.8]{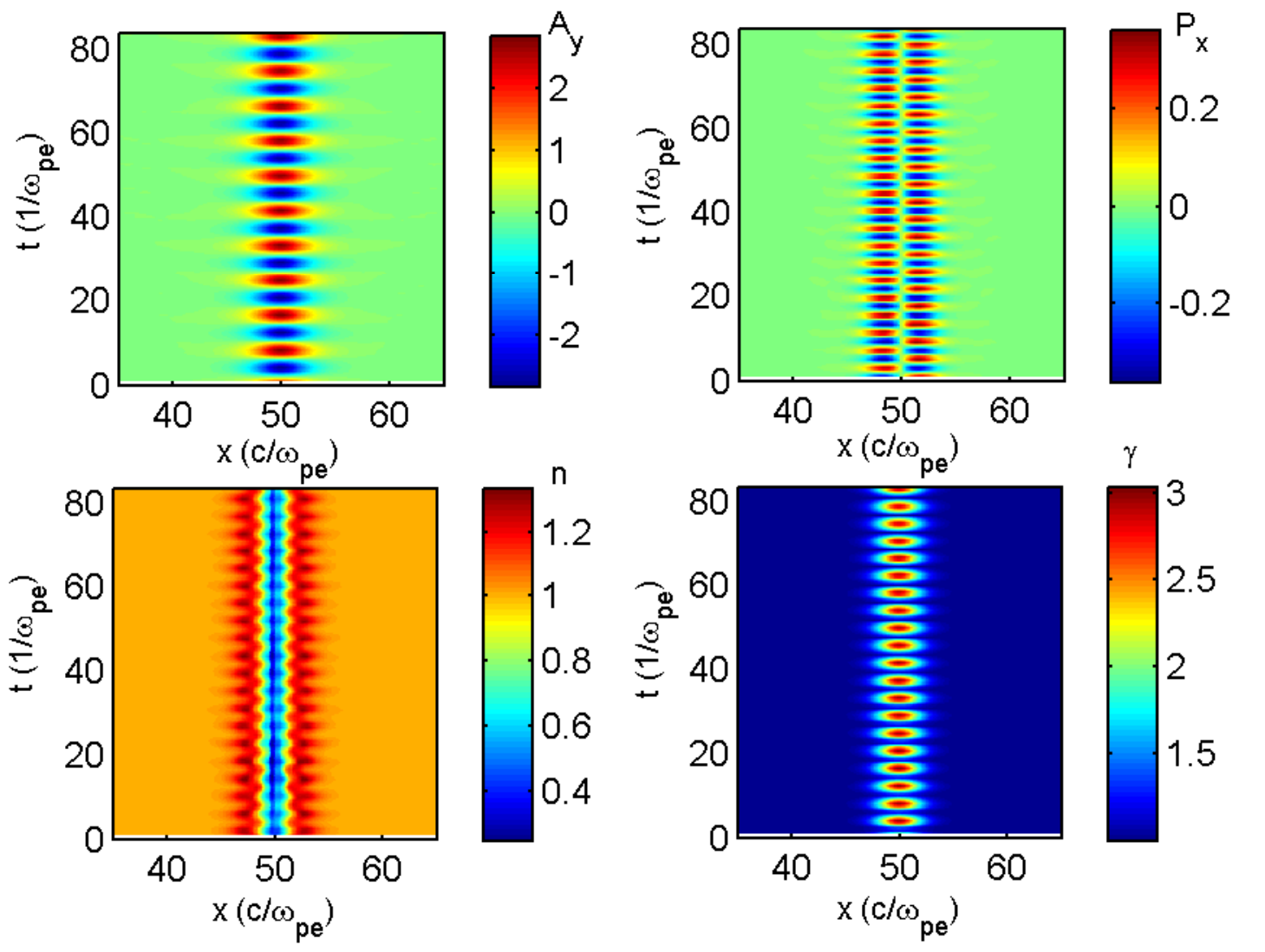}
\end{center}
\caption{\label{Fig:sim:2}\Changes{(color online) Normalized
variables of a Fluid-Maxwell simulation initialized with a solitary
wave of frequency $\Omega=0.75\omega_{pe}$.} }
\end{figure*}

The initial conditions for the numerical simulations are chosen in
accordance with the numerical solutions of Sec.
\ref{Sec:Soliton:Standing} and are then evolved with above discussed
method. In order to explore the full branch presented in Fig.
\ref{Fig:Standing:Omega}, we ran several simulations. Each one was
initialized with a solitary wave of a given frequency. For instance,
Fig. \ref{Fig:sim:2} shows a fluid-Maxwell simulation initialized
with a solitary wave of frequency $\Omega = 0.75\omega_p$,
\Changes{which is the one with highest amplitude that we simulated}.
In all cases the structures remain unchanged for several plasma
periods and they do not seem to be prone to any instability. The
amplitude and the oscillation frequency of the waves during the
simulations coincide with the values expected from the analysis of
Sec. \ref{Sec:Soliton:Standing}. \Changes{The same is true for the
frequency and wavenumber of the small oscillations in the tail}. This confirms
the integrity of the methodology and correctness of the solutions
presented in this work. As the solitary waves appear stable they are
ideal candidates for further research, for example, for the
investigation of the mutual collisions among two or more standing
structures as has been studied for the circular polarization case in
Ref. \cite{sksa,siminos2014}.

\section{\label{Sec:Conclusions}Conclusions}

We have presented novel exact \Changes{numerical} solutions of the
relativistic Maxwell-fluid model in a cold plasma with fixed ions.
One-dimensional solitary waves with linear polarization and a
breather-like behavior were computed by using a numerical method
based on a finite-difference scheme. The spatio-temporal structure
of the electromagnetic and plasma fields was presented as well as
the main properties of the waves as a function of their frequency
$\Omega$. The solitary waves exist in the frequency range
$\Omega_{min}\equiv 0.694<\Omega/\omega_{pe}<1$, with $\Omega_{min}$
the frequency value where the minimum of the plasma density
vanishes. An analysis based on the different energy contributions
showed that these localized structures consist of a plasma cavity
where a periodic exchange of energy at frequency $2\Omega$ occurs
between an electromagnetic wave and the plasma electrons.

Unlike circularly polarized waves, which are stationary, linearly
polarized waves exhibit a breather like behavior. Longitudinal
variables, like the electron momentum, oscillate with the frequency
$2\Omega$; electrons are moved inward and outward from the plasma
cavity twice per period. Besides this fundamental difference, other
features of the solitary waves are shared by both types of
polarization. For instance, the amplitude vanishes (is enhanced) and
the width increases (decreases) as the frequency approaches to
$\omega_{pe}$ ($\Omega_{min}$). For a given frequency, the amplitude
of linearly polarized waves is greater than for circular
polarization. The frequency range of existence is also broader.

\ESedit{It is well known that at the ion time scale solitons evolve to states called
postsolitons. These are slowly expanding cavities in the ion and
electron densities which trap electromagnetic energy and produce
fast ions \cite{Naumova_01,Sanchez_2011b}. However, neglecting ion motion
in our analysis is justified \emph{a posteriori}, since the typical
response time of the ions $\left(m_i/m_e\right)^{1/2}\,2\pi\,/\omega_{pe}$,
where $m_i$ the ion mass, will be much larger than the typical period of oscillation
$2\pi/\Omega$ for the solutions presented here.}

\Changes{\ESedit{In previous} fluid \cite{Wu_2013} and PIC
\cite{Macchi2012} simulations \ESedit{solitary-wave-like structures
were excited by a linearly polarized high-intensity laser pulse
incident on a plasma slab}. However, since the waves were excited
spontaneously during the complex interaction, this method has
difficulties to provide important information like the
amplitude-frequency dispersion relation or the frequency domain of
existence. In addition, it is not clear if the excited
solutions in these studies correspond to exact solitary waves. 
From this point of view, the procedure followed in this
work is advantageous, complements the information of the simulations
and opens new possibilities. For instance, our fluid simulations,
initialized with a single solitary wave, reveal that these coherent
structures persist for long time and give insight on the stability
problem.

Since part of the electromagnetic energy is trapped inside the wave
cavity, the solitary waves may play an important role in several
applications like fast ignition in inertial confinement fusion. Past
simulations, experiments and theoretical works indicate that they
are excited behind the laser pulse and they seem to be a fundamental
component of the laser-plasma interaction at high intensities. A
good understanding of some aspects, like the excitation process or
solitary wave interactions, are relevant for these applications.
These scenarios can be analyzed by preparing fluid or
particle-in-cell codes initialized with the solutions here
presented. An example of this technique applied to relativistic
solitons was given in Ref. \cite{Sanchez_11d}, where stability,
collisions, electromagnetic bursts and post-soliton evolution of
s-polarized 2-dimensional solitary waves were analyzed.}

The present analysis can be extended to find moving solitary waves
solutions with linear polarization in the relativistic fluid model.
The small amplitude limit of such solutions was presented in Ref.
\cite{Hadsievski_02}, thus suggesting that large amplitude waves may
also exist. The results of this work, that would complete our
knowledge on the organization of linearly polarized solitary waves,
will be presented elsewhere.

\acknowledgments This work was partially supported by the Ministerio
de Ciencia e Innovaci\'on of Spain (Grant No ENE2011-28489).




\begin{thebibliography}{33}%
\makeatletter
\providecommand \@ifxundefined [1]{%
 \@ifx{#1\undefined}
}%
\providecommand \@ifnum [1]{%
 \ifnum #1\expandafter \@firstoftwo
 \else \expandafter \@secondoftwo
 \fi
}%
\providecommand \@ifx [1]{%
 \ifx #1\expandafter \@firstoftwo
 \else \expandafter \@secondoftwo
 \fi
}%
\providecommand \natexlab [1]{#1}%
\providecommand \enquote  [1]{``#1''}%
\providecommand \bibnamefont  [1]{#1}%
\providecommand \bibfnamefont [1]{#1}%
\providecommand \citenamefont [1]{#1}%
\providecommand \href@noop [0]{\@secondoftwo}%
\providecommand \href [0]{\begingroup \@sanitize@url \@href}%
\providecommand \@href[1]{\@@startlink{#1}\@@href}%
\providecommand \@@href[1]{\endgroup#1\@@endlink}%
\providecommand \@sanitize@url [0]{\catcode `\\12\catcode `\$12\catcode
  `\&12\catcode `\#12\catcode `\^12\catcode `\_12\catcode `\%12\relax}%
\providecommand \@@startlink[1]{}%
\providecommand \@@endlink[0]{}%
\providecommand \url  [0]{\begingroup\@sanitize@url \@url }%
\providecommand \@url [1]{\endgroup\@href {#1}{\urlprefix }}%
\providecommand \urlprefix  [0]{URL }%
\providecommand \Eprint [0]{\href }%
\providecommand \doibase [0]{http://dx.doi.org/}%
\providecommand \selectlanguage [0]{\@gobble}%
\providecommand \bibinfo  [0]{\@secondoftwo}%
\providecommand \bibfield  [0]{\@secondoftwo}%
\providecommand \translation [1]{[#1]}%
\providecommand \BibitemOpen [0]{}%
\providecommand \bibitemStop [0]{}%
\providecommand \bibitemNoStop [0]{.\EOS\space}%
\providecommand \EOS [0]{\spacefactor3000\relax}%
\providecommand \BibitemShut  [1]{\csname bibitem#1\endcsname}%
\let\auto@bib@innerbib\@empty
\bibitem [{\citenamefont {Eliezer}\ and\ \citenamefont
  {Mima}(2008)}]{Eliezer_08}%
  \BibitemOpen
  \bibfield  {author} {\bibinfo {author} {\bibfnamefont {S.}~\bibnamefont
  {Eliezer}}\ and\ \bibinfo {author} {\bibfnamefont {K.}~\bibnamefont {Mima}},\
  }\href {http://books.google.es/books?id=inCD7lqR0K8C} {\emph {\bibinfo
  {title} {Applications of Laser-Plasma Interactions}}},\ Series in Plasma
  Physics\ (\bibinfo  {publisher} {Taylor \& Francis},\ \bibinfo {year}
  {2008})\BibitemShut {NoStop}%
\bibitem [{\citenamefont {Akhiezer}\ and\ \citenamefont
  {Polovin}(1956)}]{Akhiezer_56}%
  \BibitemOpen
  \bibfield  {author} {\bibinfo {author} {\bibfnamefont {A.}~\bibnamefont
  {Akhiezer}}\ and\ \bibinfo {author} {\bibfnamefont {R.}~\bibnamefont
  {Polovin}},\ }\href {http://www.osti.gov/scitech/servlets/purl/4361348}
  {\bibfield  {journal} {\bibinfo  {journal} {Sov. Phys. JETP}\ }\textbf
  {\bibinfo {volume} {3}},\ \bibinfo {pages} {696} (\bibinfo {year}
  {1956})}\BibitemShut {NoStop}%
\bibitem [{\citenamefont {{Kozlov}}\ \emph {et~al.}(1979)\citenamefont
  {{Kozlov}}, \citenamefont {{Litvak}},\ and\ \citenamefont
  {{Suvorov}}}]{Kozlov_79}%
  \BibitemOpen
  \bibfield  {author} {\bibinfo {author} {\bibfnamefont {V.~A.}\ \bibnamefont
  {{Kozlov}}}, \bibinfo {author} {\bibfnamefont {A.~G.}\ \bibnamefont
  {{Litvak}}}, \ and\ \bibinfo {author} {\bibfnamefont {E.~V.}\ \bibnamefont
  {{Suvorov}}},\ }\href@noop {} {\bibfield  {journal} {\bibinfo  {journal}
  {Soviet Phys. JETP}\ }\textbf {\bibinfo {volume} {49}},\ \bibinfo {pages}
  {75} (\bibinfo {year} {1979})}\BibitemShut {NoStop}%
\bibitem [{\citenamefont {Kaw}\ \emph {et~al.}(1992)\citenamefont {Kaw},
  \citenamefont {Sen},\ and\ \citenamefont {Katsouleas}}]{Kaw_92}%
  \BibitemOpen
  \bibfield  {author} {\bibinfo {author} {\bibfnamefont {P.~K.}\ \bibnamefont
  {Kaw}}, \bibinfo {author} {\bibfnamefont {A.}~\bibnamefont {Sen}}, \ and\
  \bibinfo {author} {\bibfnamefont {T.}~\bibnamefont {Katsouleas}},\ }\href
  {\doibase 10.1103/PhysRevLett.68.3172} {\bibfield  {journal} {\bibinfo
  {journal} {Phys. Rev. Lett.}\ }\textbf {\bibinfo {volume} {68}},\ \bibinfo
  {pages} {3172} (\bibinfo {year} {1992})}\BibitemShut {NoStop}%
\bibitem [{\citenamefont {{Esirkepov}}\ \emph {et~al.}(1998)\citenamefont
  {{Esirkepov}}, \citenamefont {{Kamenets}}, \citenamefont {{Bulanov}},\ and\
  \citenamefont {{Naumova}}}]{Esirkepov_1998}%
  \BibitemOpen
  \bibfield  {author} {\bibinfo {author} {\bibfnamefont {T.~Z.}\ \bibnamefont
  {{Esirkepov}}}, \bibinfo {author} {\bibfnamefont {F.~F.}\ \bibnamefont
  {{Kamenets}}}, \bibinfo {author} {\bibfnamefont {S.~V.}\ \bibnamefont
  {{Bulanov}}}, \ and\ \bibinfo {author} {\bibfnamefont {N.~M.}\ \bibnamefont
  {{Naumova}}},\ }\href {\doibase 10.1134/1.567817} {\bibfield  {journal}
  {\bibinfo  {journal} {JETP Lett.}\ }\textbf {\bibinfo {volume} {68}},\
  \bibinfo {pages} {36} (\bibinfo {year} {1998})}\BibitemShut {NoStop}%
\bibitem [{\citenamefont {{Farina}}\ and\ \citenamefont
  {{Bulanov}}(2001)}]{Farina01a}%
  \BibitemOpen
  \bibfield  {author} {\bibinfo {author} {\bibfnamefont {D.}~\bibnamefont
  {{Farina}}}\ and\ \bibinfo {author} {\bibfnamefont {S.~V.}\ \bibnamefont
  {{Bulanov}}},\ }\href {\doibase 10.1103/PhysRevLett.86.5289} {\bibfield
  {journal} {\bibinfo  {journal} {Phys. Rev. Lett.}\ }\textbf {\bibinfo
  {volume} {86}},\ \bibinfo {pages} {5289} (\bibinfo {year}
  {2001})}\BibitemShut {NoStop}%
\bibitem [{\citenamefont {{Farina}}\ and\ \citenamefont
  {{Bulanov}}(2005)}]{Farina05}%
  \BibitemOpen
  \bibfield  {author} {\bibinfo {author} {\bibfnamefont {D.}~\bibnamefont
  {{Farina}}}\ and\ \bibinfo {author} {\bibfnamefont {S.~V.}\ \bibnamefont
  {{Bulanov}}},\ }\href {\doibase 10.1088/0741-3335/47/5A/007} {\bibfield
  {journal} {\bibinfo  {journal} {Plasma Phys. Controlled Fusion}\ }\textbf
  {\bibinfo {volume} {47}},\ \bibinfo {pages} {A260000} (\bibinfo {year}
  {2005})}\BibitemShut {NoStop}%
\bibitem [{\citenamefont {{Sanchez-Arriaga}}\ \emph {et~al.}(2011)\citenamefont
  {{Sanchez-Arriaga}}, \citenamefont {{Siminos}},\ and\ \citenamefont
  {{Lefebvre}}}]{Sanchez_2011a}%
  \BibitemOpen
  \bibfield  {author} {\bibinfo {author} {\bibfnamefont {G.}~\bibnamefont
  {{Sanchez-Arriaga}}}, \bibinfo {author} {\bibfnamefont {E.}~\bibnamefont
  {{Siminos}}}, \ and\ \bibinfo {author} {\bibfnamefont {E.}~\bibnamefont
  {{Lefebvre}}},\ }\href {\doibase 10.1063/1.3624498} {\bibfield  {journal}
  {\bibinfo  {journal} {Phys. Plasmas}\ }\textbf {\bibinfo {volume} {18}},\
  \bibinfo {pages} {082304} (\bibinfo {year} {2011})}\BibitemShut {NoStop}%
\bibitem [{\citenamefont {S\'anchez-Arriaga}\ \emph {et~al.}(2011)\citenamefont
  {S\'anchez-Arriaga}, \citenamefont {Siminos},\ and\ \citenamefont
  {Lefebvre}}]{Sanchez_2011b}%
  \BibitemOpen
  \bibfield  {author} {\bibinfo {author} {\bibfnamefont {G.}~\bibnamefont
  {S\'anchez-Arriaga}}, \bibinfo {author} {\bibfnamefont {E.}~\bibnamefont
  {Siminos}}, \ and\ \bibinfo {author} {\bibfnamefont {E.}~\bibnamefont
  {Lefebvre}},\ }\href {http://stacks.iop.org/0741-3335/53/i=4/a=045011}
  {\bibfield  {journal} {\bibinfo  {journal} {Plasma Physics and Controlled
  Fusion}\ }\textbf {\bibinfo {volume} {53}},\ \bibinfo {pages} {045011}
  (\bibinfo {year} {2011})}\BibitemShut {NoStop}%
\bibitem [{\citenamefont {{Had{\v z}ievski}}\ \emph {et~al.}(2002)\citenamefont
  {{Had{\v z}ievski}}, \citenamefont {{Jovanovi{\'c}}}, \citenamefont {{{\v
  S}kori{\'c}}},\ and\ \citenamefont {{Mima}}}]{Hadsievski_02}%
  \BibitemOpen
  \bibfield  {author} {\bibinfo {author} {\bibfnamefont {L.}~\bibnamefont
  {{Had{\v z}ievski}}}, \bibinfo {author} {\bibfnamefont {M.~S.}\ \bibnamefont
  {{Jovanovi{\'c}}}}, \bibinfo {author} {\bibfnamefont {M.~M.}\ \bibnamefont
  {{{\v S}kori{\'c}}}}, \ and\ \bibinfo {author} {\bibfnamefont
  {K.}~\bibnamefont {{Mima}}},\ }\href {\doibase 10.1063/1.1476665} {\bibfield
  {journal} {\bibinfo  {journal} {Phys. Plasmas}\ }\textbf {\bibinfo {volume}
  {9}},\ \bibinfo {pages} {2569} (\bibinfo {year} {2002})}\BibitemShut
  {NoStop}%
\bibitem [{\citenamefont {Mancic}\ \emph {et~al.}(2006)\citenamefont {Mancic},
  \citenamefont {Had{\v z}ievski},\ and\ \citenamefont {{\v
  S}kori{\'c}}}]{Mancic_2006}%
  \BibitemOpen
  \bibfield  {author} {\bibinfo {author} {\bibfnamefont {A.}~\bibnamefont
  {Mancic}}, \bibinfo {author} {\bibfnamefont {L.}~\bibnamefont {Had{\v
  z}ievski}}, \ and\ \bibinfo {author} {\bibfnamefont {M.~M.}\ \bibnamefont
  {{\v S}kori{\'c}}},\ }\href {\doibase http://dx.doi.org/10.1063/1.2203606}
  {\bibfield  {journal} {\bibinfo  {journal} {Phys. Plasmas}\ }\textbf
  {\bibinfo {volume} {13}},\ \bibinfo {eid} {052309} (\bibinfo {year}
  {2006})}\BibitemShut {NoStop}%
\bibitem [{\citenamefont {{Saxena}}\ \emph {et~al.}(2009)\citenamefont
  {{Saxena}}, \citenamefont {{Sen}},\ and\ \citenamefont {{Kaw}}}]{Saxena_09}%
  \BibitemOpen
  \bibfield  {author} {\bibinfo {author} {\bibfnamefont {V.}~\bibnamefont
  {{Saxena}}}, \bibinfo {author} {\bibfnamefont {A.}~\bibnamefont {{Sen}}}, \
  and\ \bibinfo {author} {\bibfnamefont {P.}~\bibnamefont {{Kaw}}},\ }\href
  {\doibase 10.1103/PhysRevE.80.016406} {\bibfield  {journal} {\bibinfo
  {journal} {\pre}\ }\textbf {\bibinfo {volume} {80}},\ \bibinfo {eid} {016406}
  (\bibinfo {year} {2009})}\BibitemShut {NoStop}%
\bibitem [{\citenamefont {{S{\'a}nchez-Arriaga}}\ and\ \citenamefont
  {{Lefebvre}}(2011{\natexlab{a}})}]{Sanchez_11}%
  \BibitemOpen
  \bibfield  {author} {\bibinfo {author} {\bibfnamefont {G.}~\bibnamefont
  {{S{\'a}nchez-Arriaga}}}\ and\ \bibinfo {author} {\bibfnamefont
  {E.}~\bibnamefont {{Lefebvre}}},\ }\href {\doibase
  10.1103/PhysRevE.84.036403} {\bibfield  {journal} {\bibinfo  {journal}
  {\pre}\ }\textbf {\bibinfo {volume} {84}},\ \bibinfo {eid} {036403} (\bibinfo
  {year} {2011}{\natexlab{a}})}\BibitemShut {NoStop}%
\bibitem [{\citenamefont {{Bulanov}}\ \emph {et~al.}(1992)\citenamefont
  {{Bulanov}}, \citenamefont {{Inovenkov}}, \citenamefont {{Kirsanov}},
  \citenamefont {{Naumova}},\ and\ \citenamefont {{Sakharov}}}]{Bulanov_92}%
  \BibitemOpen
  \bibfield  {author} {\bibinfo {author} {\bibfnamefont {S.~V.}\ \bibnamefont
  {{Bulanov}}}, \bibinfo {author} {\bibfnamefont {I.~N.}\ \bibnamefont
  {{Inovenkov}}}, \bibinfo {author} {\bibfnamefont {V.~I.}\ \bibnamefont
  {{Kirsanov}}}, \bibinfo {author} {\bibfnamefont {N.~M.}\ \bibnamefont
  {{Naumova}}}, \ and\ \bibinfo {author} {\bibfnamefont {A.~S.}\ \bibnamefont
  {{Sakharov}}},\ }\href {\doibase 10.1063/1.860046} {\bibfield  {journal}
  {\bibinfo  {journal} {Physics of Fluids B}\ }\textbf {\bibinfo {volume}
  {4}},\ \bibinfo {pages} {1935} (\bibinfo {year} {1992})}\BibitemShut
  {NoStop}%
\bibitem [{\citenamefont {Bulanov}\ \emph {et~al.}(1999)\citenamefont
  {Bulanov}, \citenamefont {Esirkepov}, \citenamefont {Naumova}, \citenamefont
  {Pegoraro},\ and\ \citenamefont {Vshivkov}}]{Bulanov_99}%
  \BibitemOpen
  \bibfield  {author} {\bibinfo {author} {\bibfnamefont {S.~V.}\ \bibnamefont
  {Bulanov}}, \bibinfo {author} {\bibfnamefont {T.~Z.}\ \bibnamefont
  {Esirkepov}}, \bibinfo {author} {\bibfnamefont {N.~M.}\ \bibnamefont
  {Naumova}}, \bibinfo {author} {\bibfnamefont {F.}~\bibnamefont {Pegoraro}}, \
  and\ \bibinfo {author} {\bibfnamefont {V.~A.}\ \bibnamefont {Vshivkov}},\
  }\href {\doibase 10.1103/PhysRevLett.82.3440} {\bibfield  {journal} {\bibinfo
   {journal} {Phys. Rev. Lett.}\ }\textbf {\bibinfo {volume} {82}},\ \bibinfo
  {pages} {3440} (\bibinfo {year} {1999})}\BibitemShut {NoStop}%
\bibitem [{\citenamefont {Esirkepov}\ \emph {et~al.}(2002)\citenamefont
  {Esirkepov}, \citenamefont {Nishihara}, \citenamefont {Bulanov},\ and\
  \citenamefont {Pegoraro}}]{Esirkepov_02}%
  \BibitemOpen
  \bibfield  {author} {\bibinfo {author} {\bibfnamefont {T.}~\bibnamefont
  {Esirkepov}}, \bibinfo {author} {\bibfnamefont {K.}~\bibnamefont
  {Nishihara}}, \bibinfo {author} {\bibfnamefont {S.~V.}\ \bibnamefont
  {Bulanov}}, \ and\ \bibinfo {author} {\bibfnamefont {F.}~\bibnamefont
  {Pegoraro}},\ }\href {\doibase 10.1103/PhysRevLett.89.275002} {\bibfield
  {journal} {\bibinfo  {journal} {Phys. Rev. Lett.}\ }\textbf {\bibinfo
  {volume} {89}},\ \bibinfo {pages} {275002} (\bibinfo {year}
  {2002})}\BibitemShut {NoStop}%
\bibitem [{\citenamefont {Borghesi}\ \emph {et~al.}(2002)\citenamefont
  {Borghesi}, \citenamefont {Bulanov}, \citenamefont {Campbell}, \citenamefont
  {Clarke}, \citenamefont {Esirkepov}, \citenamefont {Galimberti},
  \citenamefont {Gizzi}, \citenamefont {MacKinnon}, \citenamefont {Naumova},
  \citenamefont {Pegoraro}, \citenamefont {Ruhl}, \citenamefont {Schiavi},\
  and\ \citenamefont {Willi}}]{Borghesi_02}%
  \BibitemOpen
  \bibfield  {author} {\bibinfo {author} {\bibfnamefont {M.}~\bibnamefont
  {Borghesi}}, \bibinfo {author} {\bibfnamefont {S.}~\bibnamefont {Bulanov}},
  \bibinfo {author} {\bibfnamefont {D.~H.}\ \bibnamefont {Campbell}}, \bibinfo
  {author} {\bibfnamefont {R.~J.}\ \bibnamefont {Clarke}}, \bibinfo {author}
  {\bibfnamefont {T.~Z.}\ \bibnamefont {Esirkepov}}, \bibinfo {author}
  {\bibfnamefont {M.}~\bibnamefont {Galimberti}}, \bibinfo {author}
  {\bibfnamefont {L.~A.}\ \bibnamefont {Gizzi}}, \bibinfo {author}
  {\bibfnamefont {A.~J.}\ \bibnamefont {MacKinnon}}, \bibinfo {author}
  {\bibfnamefont {N.~M.}\ \bibnamefont {Naumova}}, \bibinfo {author}
  {\bibfnamefont {F.}~\bibnamefont {Pegoraro}}, \bibinfo {author}
  {\bibfnamefont {H.}~\bibnamefont {Ruhl}}, \bibinfo {author} {\bibfnamefont
  {A.}~\bibnamefont {Schiavi}}, \ and\ \bibinfo {author} {\bibfnamefont
  {O.}~\bibnamefont {Willi}},\ }\href {\doibase 10.1103/PhysRevLett.88.135002}
  {\bibfield  {journal} {\bibinfo  {journal} {Phys. Rev. Lett.}\ }\textbf
  {\bibinfo {volume} {88}},\ \bibinfo {pages} {135002} (\bibinfo {year}
  {2002})}\BibitemShut {NoStop}%
\bibitem [{\citenamefont {Chen}\ \emph {et~al.}(2007)\citenamefont {Chen},
  \citenamefont {Kotaki}, \citenamefont {Nakajima}, \citenamefont {Koga},
  \citenamefont {Bulanov}, \citenamefont {Tajima}, \citenamefont {Gu},
  \citenamefont {Peng}, \citenamefont {Wang}, \citenamefont {Wen},
  \citenamefont {Liu}, \citenamefont {Jiao}, \citenamefont {Zhang},
  \citenamefont {Huang}, \citenamefont {Guo}, \citenamefont {Zhou},
  \citenamefont {Hua}, \citenamefont {An}, \citenamefont {Tang},\ and\
  \citenamefont {Lin}}]{Chen_07}%
  \BibitemOpen
  \bibfield  {author} {\bibinfo {author} {\bibfnamefont {L.~M.}\ \bibnamefont
  {Chen}}, \bibinfo {author} {\bibfnamefont {H.}~\bibnamefont {Kotaki}},
  \bibinfo {author} {\bibfnamefont {K.}~\bibnamefont {Nakajima}}, \bibinfo
  {author} {\bibfnamefont {J.}~\bibnamefont {Koga}}, \bibinfo {author}
  {\bibfnamefont {S.~V.}\ \bibnamefont {Bulanov}}, \bibinfo {author}
  {\bibfnamefont {T.}~\bibnamefont {Tajima}}, \bibinfo {author} {\bibfnamefont
  {Y.~Q.}\ \bibnamefont {Gu}}, \bibinfo {author} {\bibfnamefont {H.~S.}\
  \bibnamefont {Peng}}, \bibinfo {author} {\bibfnamefont {X.~X.}\ \bibnamefont
  {Wang}}, \bibinfo {author} {\bibfnamefont {T.~S.}\ \bibnamefont {Wen}},
  \bibinfo {author} {\bibfnamefont {H.~J.}\ \bibnamefont {Liu}}, \bibinfo
  {author} {\bibfnamefont {C.~Y.}\ \bibnamefont {Jiao}}, \bibinfo {author}
  {\bibfnamefont {C.~G.}\ \bibnamefont {Zhang}}, \bibinfo {author}
  {\bibfnamefont {X.~J.}\ \bibnamefont {Huang}}, \bibinfo {author}
  {\bibfnamefont {Y.}~\bibnamefont {Guo}}, \bibinfo {author} {\bibfnamefont
  {K.~N.}\ \bibnamefont {Zhou}}, \bibinfo {author} {\bibfnamefont {J.~F.}\
  \bibnamefont {Hua}}, \bibinfo {author} {\bibfnamefont {W.~M.}\ \bibnamefont
  {An}}, \bibinfo {author} {\bibfnamefont {C.~X.}\ \bibnamefont {Tang}}, \ and\
  \bibinfo {author} {\bibfnamefont {Y.~Z.}\ \bibnamefont {Lin}},\ }\href
  {\doibase http://dx.doi.org/10.1063/1.2720374} {\bibfield  {journal}
  {\bibinfo  {journal} {Phys. Plasmas}\ }\textbf {\bibinfo {volume} {14}},\
  \bibinfo {eid} {040703} (\bibinfo {year} {2007})}\BibitemShut {NoStop}%
\bibitem [{\citenamefont {Pirozhkov}\ \emph {et~al.}(2007)\citenamefont
  {Pirozhkov}, \citenamefont {Ma}, \citenamefont {Kando}, \citenamefont
  {Esirkepov}, \citenamefont {Fukuda}, \citenamefont {Chen}, \citenamefont
  {Daito}, \citenamefont {Ogura}, \citenamefont {Homma}, \citenamefont
  {Hayashi}, \citenamefont {Kotaki}, \citenamefont {Sagisaka}, \citenamefont
  {Mori}, \citenamefont {Koga}, \citenamefont {Kawachi}, \citenamefont {Daido},
  \citenamefont {Bulanov}, \citenamefont {Kimura}, \citenamefont {Kato},\ and\
  \citenamefont {Tajima}}]{Pirozhkov_07}%
  \BibitemOpen
  \bibfield  {author} {\bibinfo {author} {\bibfnamefont {A.~S.}\ \bibnamefont
  {Pirozhkov}}, \bibinfo {author} {\bibfnamefont {J.}~\bibnamefont {Ma}},
  \bibinfo {author} {\bibfnamefont {M.}~\bibnamefont {Kando}}, \bibinfo
  {author} {\bibfnamefont {T.~Z.}\ \bibnamefont {Esirkepov}}, \bibinfo {author}
  {\bibfnamefont {Y.}~\bibnamefont {Fukuda}}, \bibinfo {author} {\bibfnamefont
  {L.-M.}\ \bibnamefont {Chen}}, \bibinfo {author} {\bibfnamefont
  {I.}~\bibnamefont {Daito}}, \bibinfo {author} {\bibfnamefont
  {K.}~\bibnamefont {Ogura}}, \bibinfo {author} {\bibfnamefont
  {T.}~\bibnamefont {Homma}}, \bibinfo {author} {\bibfnamefont
  {Y.}~\bibnamefont {Hayashi}}, \bibinfo {author} {\bibfnamefont
  {H.}~\bibnamefont {Kotaki}}, \bibinfo {author} {\bibfnamefont
  {A.}~\bibnamefont {Sagisaka}}, \bibinfo {author} {\bibfnamefont
  {M.}~\bibnamefont {Mori}}, \bibinfo {author} {\bibfnamefont {J.~K.}\
  \bibnamefont {Koga}}, \bibinfo {author} {\bibfnamefont {T.}~\bibnamefont
  {Kawachi}}, \bibinfo {author} {\bibfnamefont {H.}~\bibnamefont {Daido}},
  \bibinfo {author} {\bibfnamefont {S.~V.}\ \bibnamefont {Bulanov}}, \bibinfo
  {author} {\bibfnamefont {T.}~\bibnamefont {Kimura}}, \bibinfo {author}
  {\bibfnamefont {Y.}~\bibnamefont {Kato}}, \ and\ \bibinfo {author}
  {\bibfnamefont {T.}~\bibnamefont {Tajima}},\ }\href {\doibase
  http://dx.doi.org/10.1063/1.2816443} {\bibfield  {journal} {\bibinfo
  {journal} {Phys. Plasmas}\ }\textbf {\bibinfo {volume} {14}},\ \bibinfo {eid}
  {123106} (\bibinfo {year} {2007})}\BibitemShut {NoStop}%
\bibitem [{\citenamefont {Sarri}\ \emph {et~al.}(2010)\citenamefont {Sarri},
  \citenamefont {Singh}, \citenamefont {Davies}, \citenamefont {Fiuza},
  \citenamefont {Lancaster}, \citenamefont {Clark}, \citenamefont {Hassan},
  \citenamefont {Jiang}, \citenamefont {Kageiwa}, \citenamefont {Lopes},
  \citenamefont {Rehman}, \citenamefont {Russo}, \citenamefont {Scott},
  \citenamefont {Tanimoto}, \citenamefont {Najmudin}, \citenamefont {Tanaka},
  \citenamefont {Tatarakis}, \citenamefont {Borghesi},\ and\ \citenamefont
  {Norreys}}]{Sarri_10}%
  \BibitemOpen
  \bibfield  {author} {\bibinfo {author} {\bibfnamefont {G.}~\bibnamefont
  {Sarri}}, \bibinfo {author} {\bibfnamefont {D.~K.}\ \bibnamefont {Singh}},
  \bibinfo {author} {\bibfnamefont {J.~R.}\ \bibnamefont {Davies}}, \bibinfo
  {author} {\bibfnamefont {F.}~\bibnamefont {Fiuza}}, \bibinfo {author}
  {\bibfnamefont {K.~L.}\ \bibnamefont {Lancaster}}, \bibinfo {author}
  {\bibfnamefont {E.~L.}\ \bibnamefont {Clark}}, \bibinfo {author}
  {\bibfnamefont {S.}~\bibnamefont {Hassan}}, \bibinfo {author} {\bibfnamefont
  {J.}~\bibnamefont {Jiang}}, \bibinfo {author} {\bibfnamefont
  {N.}~\bibnamefont {Kageiwa}}, \bibinfo {author} {\bibfnamefont
  {N.}~\bibnamefont {Lopes}}, \bibinfo {author} {\bibfnamefont
  {A.}~\bibnamefont {Rehman}}, \bibinfo {author} {\bibfnamefont
  {C.}~\bibnamefont {Russo}}, \bibinfo {author} {\bibfnamefont {R.~H.~H.}\
  \bibnamefont {Scott}}, \bibinfo {author} {\bibfnamefont {T.}~\bibnamefont
  {Tanimoto}}, \bibinfo {author} {\bibfnamefont {Z.}~\bibnamefont {Najmudin}},
  \bibinfo {author} {\bibfnamefont {K.~A.}\ \bibnamefont {Tanaka}}, \bibinfo
  {author} {\bibfnamefont {M.}~\bibnamefont {Tatarakis}}, \bibinfo {author}
  {\bibfnamefont {M.}~\bibnamefont {Borghesi}}, \ and\ \bibinfo {author}
  {\bibfnamefont {P.~A.}\ \bibnamefont {Norreys}},\ }\href {\doibase
  10.1103/PhysRevLett.105.175007} {\bibfield  {journal} {\bibinfo  {journal}
  {Phys. Rev. Lett.}\ }\textbf {\bibinfo {volume} {105}},\ \bibinfo {pages}
  {175007} (\bibinfo {year} {2010})}\BibitemShut {NoStop}%
\bibitem [{\citenamefont {{Sylla}}\ \emph {et~al.}(2012)\citenamefont
  {{Sylla}}, \citenamefont {{Flacco}}, \citenamefont {{Kahaly}}, \citenamefont
  {{Veltcheva}}, \citenamefont {{Lifschitz}}, \citenamefont
  {{Sanchez-Arriaga}}, \citenamefont {{Lefebvre}},\ and\ \citenamefont
  {{Malka}}}]{Sylla_12}%
  \BibitemOpen
  \bibfield  {author} {\bibinfo {author} {\bibfnamefont {F.}~\bibnamefont
  {{Sylla}}}, \bibinfo {author} {\bibfnamefont {A.}~\bibnamefont {{Flacco}}},
  \bibinfo {author} {\bibfnamefont {S.}~\bibnamefont {{Kahaly}}}, \bibinfo
  {author} {\bibfnamefont {M.}~\bibnamefont {{Veltcheva}}}, \bibinfo {author}
  {\bibfnamefont {A.}~\bibnamefont {{Lifschitz}}}, \bibinfo {author}
  {\bibfnamefont {G.}~\bibnamefont {{Sanchez-Arriaga}}}, \bibinfo {author}
  {\bibfnamefont {E.}~\bibnamefont {{Lefebvre}}}, \ and\ \bibinfo {author}
  {\bibfnamefont {V.}~\bibnamefont {{Malka}}},\ }\href {\doibase
  10.1103/PhysRevLett.108.115003} {\bibfield  {journal} {\bibinfo  {journal}
  {Phys. Rev. Lett.}\ }\textbf {\bibinfo {volume} {108}},\ \bibinfo {eid}
  {115003} (\bibinfo {year} {2012})},\ \Eprint {http://arxiv.org/abs/1107.2560}
  {arXiv:1107.2560 [physics.plasm-ph]} \BibitemShut {NoStop}%
\bibitem [{\citenamefont {{Wu}}\ \emph {et~al.}(2013)\citenamefont {{Wu}},
  \citenamefont {{Zheng}}, \citenamefont {{Yan}}, \citenamefont {{Yu}},\ and\
  \citenamefont {{He}}}]{Wu_2013}%
  \BibitemOpen
  \bibfield  {author} {\bibinfo {author} {\bibfnamefont {D.}~\bibnamefont
  {{Wu}}}, \bibinfo {author} {\bibfnamefont {C.~Y.}\ \bibnamefont {{Zheng}}},
  \bibinfo {author} {\bibfnamefont {X.~Q.}\ \bibnamefont {{Yan}}}, \bibinfo
  {author} {\bibfnamefont {M.~Y.}\ \bibnamefont {{Yu}}}, \ and\ \bibinfo
  {author} {\bibfnamefont {X.~T.}\ \bibnamefont {{He}}},\ }\href {\doibase
  10.1063/1.4794197} {\bibfield  {journal} {\bibinfo  {journal} {Phys.
  Plasmas}\ }\textbf {\bibinfo {volume} {20}},\ \bibinfo {pages} {033101}
  (\bibinfo {year} {2013})},\ \Eprint {http://arxiv.org/abs/1212.0324}
  {arXiv:1212.0324 [physics.plasm-ph]} \BibitemShut {NoStop}%
\bibitem [{\citenamefont {Macchi}\ \emph {et~al.}(2012)\citenamefont {Macchi},
  \citenamefont {Nindrayog},\ and\ \citenamefont {Pegoraro}}]{Macchi2012}%
  \BibitemOpen
  \bibfield  {author} {\bibinfo {author} {\bibfnamefont {A.}~\bibnamefont
  {Macchi}}, \bibinfo {author} {\bibfnamefont {A.~S.}\ \bibnamefont
  {Nindrayog}}, \ and\ \bibinfo {author} {\bibfnamefont {F.}~\bibnamefont
  {Pegoraro}},\ }\href {\doibase 10.1103/PhysRevE.85.046402} {\bibfield
  {journal} {\bibinfo  {journal} {Phys. Rev. E}\ }\textbf {\bibinfo {volume}
  {85}},\ \bibinfo {pages} {046402} (\bibinfo {year} {2012})}\BibitemShut
  {NoStop}%
\bibitem [{\citenamefont {Poornakala}\ \emph {et~al.}(2002)\citenamefont
  {Poornakala}, \citenamefont {Das}, \citenamefont {Kaw}, \citenamefont {Sen},
  \citenamefont {Sheng}, \citenamefont {Sentoku}, \citenamefont {Mima},\ and\
  \citenamefont {Nishikawa}}]{Poornakala_2002}%
  \BibitemOpen
  \bibfield  {author} {\bibinfo {author} {\bibfnamefont {S.}~\bibnamefont
  {Poornakala}}, \bibinfo {author} {\bibfnamefont {A.}~\bibnamefont {Das}},
  \bibinfo {author} {\bibfnamefont {P.~K.}\ \bibnamefont {Kaw}}, \bibinfo
  {author} {\bibfnamefont {A.}~\bibnamefont {Sen}}, \bibinfo {author}
  {\bibfnamefont {Z.~M.}\ \bibnamefont {Sheng}}, \bibinfo {author}
  {\bibfnamefont {Y.}~\bibnamefont {Sentoku}}, \bibinfo {author} {\bibfnamefont
  {K.}~\bibnamefont {Mima}}, \ and\ \bibinfo {author} {\bibfnamefont
  {K.}~\bibnamefont {Nishikawa}},\ }\href@noop {} {\bibfield  {journal}
  {\bibinfo  {journal} {Physics of Plasmas (1994-present)}\ }\textbf {\bibinfo
  {volume} {9}} (\bibinfo {year} {2002})}\BibitemShut {NoStop}%
\bibitem [{\citenamefont {{Lehmann}}\ and\ \citenamefont
  {{Spatschek}}(2011)}]{Lehmann_11}%
  \BibitemOpen
  \bibfield  {author} {\bibinfo {author} {\bibfnamefont {G.}~\bibnamefont
  {{Lehmann}}}\ and\ \bibinfo {author} {\bibfnamefont {K.~H.}\ \bibnamefont
  {{Spatschek}}},\ }\href@noop {} {\bibfield  {journal} {\bibinfo  {journal}
  {Physical Review E}\ }\textbf {\bibinfo {volume} {83}},\ \bibinfo {pages}
  {036401} (\bibinfo {year} {2011})}\BibitemShut {NoStop}%
\bibitem [{\citenamefont {Pesch}\ and\ \citenamefont {Kull}(2007)}]{Pesch_07}%
  \BibitemOpen
  \bibfield  {author} {\bibinfo {author} {\bibfnamefont {T.~C.}\ \bibnamefont
  {Pesch}}\ and\ \bibinfo {author} {\bibfnamefont {H.-J.}\ \bibnamefont
  {Kull}},\ }\href {\doibase http://dx.doi.org/10.1063/1.2760209} {\bibfield
  {journal} {\bibinfo  {journal} {Phys. Plasmas}\ }\textbf {\bibinfo {volume}
  {14}},\ \bibinfo {eid} {083103} (\bibinfo {year} {2007})}\BibitemShut
  {NoStop}%
\bibitem [{\citenamefont {Champneys}(1998)}]{Champneys97}%
  \BibitemOpen
  \bibfield  {author} {\bibinfo {author} {\bibfnamefont {A.}~\bibnamefont
  {Champneys}},\ }\href {\doibase
  http://dx.doi.org/10.1016/S0167-2789(97)00209-1} {\bibfield  {journal}
  {\bibinfo  {journal} {Physica D}\ }\textbf {\bibinfo {volume} {112}},\
  \bibinfo {pages} {158} (\bibinfo {year} {1998})}\BibitemShut {NoStop}%
\bibitem [{\citenamefont {{Boris}}\ \emph {et~al.}(1993)\citenamefont
  {{Boris}}, \citenamefont {{Landsberg}}, \citenamefont {{Oran}},\ and\
  \citenamefont {{Gardner}}}]{lcpfct}%
  \BibitemOpen
  \bibfield  {author} {\bibinfo {author} {\bibfnamefont {J.~P.}\ \bibnamefont
  {{Boris}}}, \bibinfo {author} {\bibfnamefont {A.~M.}\ \bibnamefont
  {{Landsberg}}}, \bibinfo {author} {\bibfnamefont {E.~S.}\ \bibnamefont
  {{Oran}}}, \ and\ \bibinfo {author} {\bibfnamefont {J.~H.}\ \bibnamefont
  {{Gardner}}},\ }\href@noop {} {\emph {\bibinfo {title} {{LCPFCT-A
  flux-corrected transport algorithm for solving generalized continuity
  equations}}}},\ \bibinfo {type} {Tech. Rep.}\ \bibinfo {number}
  {NRL/MR/6410--93-7192}\ (\bibinfo {year} {1993})\BibitemShut {NoStop}%
\bibitem [{\citenamefont {{Boris}}\ and\ \citenamefont {{Book}}(1976)}]{boris}%
  \BibitemOpen
  \bibfield  {author} {\bibinfo {author} {\bibfnamefont {J.~P.}\ \bibnamefont
  {{Boris}}}\ and\ \bibinfo {author} {\bibfnamefont {D.~L.}\ \bibnamefont
  {{Book}}},\ }\href@noop {} {\bibfield  {journal} {\bibinfo  {journal}
  {Methods Comput. Phys.}\ }\textbf {\bibinfo {volume} {16}},\ \bibinfo {pages}
  {85} (\bibinfo {year} {1976})}\BibitemShut {NoStop}%
\bibitem [{\citenamefont {Siminos}\ \emph {et~al.}(2014)\citenamefont
  {Siminos}, \citenamefont {S\'anchez-Arriaga}, \citenamefont {Saxena},\ and\
  \citenamefont {Kourakis}}]{siminos2014}%
  \BibitemOpen
  \bibfield  {author} {\bibinfo {author} {\bibfnamefont {E.}~\bibnamefont
  {Siminos}}, \bibinfo {author} {\bibfnamefont {G.}~\bibnamefont
  {S\'anchez-Arriaga}}, \bibinfo {author} {\bibfnamefont {V.}~\bibnamefont
  {Saxena}}, \ and\ \bibinfo {author} {\bibfnamefont {I.}~\bibnamefont
  {Kourakis}},\ }\href {\doibase 10.1103/PhysRevE.90.063104} {\bibfield
  {journal} {\bibinfo  {journal} {Phys. Rev. E}\ }\textbf {\bibinfo {volume}
  {90}},\ \bibinfo {pages} {063104} (\bibinfo {year} {2014})}\BibitemShut
  {NoStop}%
\bibitem [{\citenamefont {{Saxena}}\ \emph {et~al.}(2013)\citenamefont
  {{Saxena}}, \citenamefont {{Kourakis}}, \citenamefont {{S\'anchez-Arriaga}},\
  and\ \citenamefont {{Siminos}}}]{sksa}%
  \BibitemOpen
  \bibfield  {author} {\bibinfo {author} {\bibfnamefont {V.}~\bibnamefont
  {{Saxena}}}, \bibinfo {author} {\bibfnamefont {I.}~\bibnamefont
  {{Kourakis}}}, \bibinfo {author} {\bibfnamefont {G.}~\bibnamefont
  {{S\'anchez-Arriaga}}}, \ and\ \bibinfo {author} {\bibfnamefont
  {E.}~\bibnamefont {{Siminos}}},\ }\href@noop {} {\bibfield  {journal}
  {\bibinfo  {journal} {Phys. Lett. A}\ }\textbf {\bibinfo {volume} {377}},\
  \bibinfo {pages} {473} (\bibinfo {year} {2013})}\BibitemShut {NoStop}%
\bibitem [{\citenamefont {Naumova}\ \emph {et~al.}(2001)\citenamefont
  {Naumova}, \citenamefont {Bulanov}, \citenamefont {Esirkepov}, \citenamefont
  {Farina}, \citenamefont {Nishihara}, \citenamefont {Pegoraro}, \citenamefont
  {Ruhl},\ and\ \citenamefont {Sakharov}}]{Naumova_01}%
  \BibitemOpen
  \bibfield  {author} {\bibinfo {author} {\bibfnamefont {N.}~\bibnamefont
  {Naumova}}, \bibinfo {author} {\bibfnamefont {S.}~\bibnamefont {Bulanov}},
  \bibinfo {author} {\bibfnamefont {T.}~\bibnamefont {Esirkepov}}, \bibinfo
  {author} {\bibfnamefont {D.}~\bibnamefont {Farina}}, \bibinfo {author}
  {\bibfnamefont {K.}~\bibnamefont {Nishihara}}, \bibinfo {author}
  {\bibfnamefont {F.}~\bibnamefont {Pegoraro}}, \bibinfo {author}
  {\bibfnamefont {H.}~\bibnamefont {Ruhl}}, \ and\ \bibinfo {author}
  {\bibfnamefont {A.}~\bibnamefont {Sakharov}},\ }\href {\doibase
  10.1103/PhysRevLett.87.185004} {\bibfield  {journal} {\bibinfo  {journal}
  {Phys. Rev. Lett.}\ }\textbf {\bibinfo {volume} {87}},\ \bibinfo {pages}
  {185004} (\bibinfo {year} {2001})}\BibitemShut {NoStop}%
\bibitem [{\citenamefont {{S{\'a}nchez-Arriaga}}\ and\ \citenamefont
  {{Lefebvre}}(2011{\natexlab{b}})}]{Sanchez_11d}%
  \BibitemOpen
  \bibfield  {author} {\bibinfo {author} {\bibfnamefont {G.}~\bibnamefont
  {{S{\'a}nchez-Arriaga}}}\ and\ \bibinfo {author} {\bibfnamefont
  {E.}~\bibnamefont {{Lefebvre}}},\ }\href@noop {} {\bibfield  {journal}
  {\bibinfo  {journal} {\pre}\ }\textbf {\bibinfo {volume} {84}},\ \bibinfo
  {eid} {036404} (\bibinfo {year} {2011}{\natexlab{b}})}\BibitemShut {NoStop}%
\end{thebibliography}
%

\end{document}